\documentclass[
reprint,
superscriptaddress, 
aps,
pra, 
showpacs, 
showkeys,
notitlepage,
amsmath, 
amssymb, 
]{revtex4-1}

\usepackage{graphicx}
\usepackage{bm}
\usepackage{color}
\usepackage[svgnames]{xcolor}
\usepackage{hyperref}
\usepackage[mathlines]{lineno}
\usepackage{epstopdf}

\newcommand*\diff{\mathop{}\!\mathrm{d}}

\renewcommand{\Re}{\operatorname{Re}}

\providecommand*{\pderiv}[3][]{ \frac{\partial^{#1}#2}{\partial #3^{#1}} }

\def\NFA{NaFeAs}
\def\NFCA{NaFe$_{1-x}$Cu$_{x}$As}
\def\NFNA{NaFe$_{1-x}$Ni$_{x}$As}
\def\LFA{LiFeAs}
\def\NFCoA{NaFe$_{1-x}$Co$_{x}$As}
\def\BFCoA{BaFe$_{2-x}$Co$_{x}$As$_{2}$}
\def\BFNA{BaFe$_{2-x}$Ni$_{x}$As$_{2}$}
\def\BKFA{Ba$_{1-x}$K$_{x}$Fe$_{2}$As$_{2}$}

\def\MuSR{$\mu$SR}

\def\muN{$\mu_{\textrm{N}}$}
\def\muB{$\mu_{\textrm{B}}$}

\def\TC{$T_\textrm{C}$}
\def\TN{$T_\textrm{N}$}
\def\TS{$T_\textrm{S}$}

\hypersetup{
	pdfpagemode={UseOutlines},
	pdfpagelayout={SinglePage},
	bookmarksopen=true,
	bookmarksnumbered = true,
	pdfstartview={FitV},
	colorlinks = true,
	linkcolor = DarkBlue,
	citecolor = DarkBlue,
	urlcolor = DarkBlue
}

\begin{document}
	
	\title{Disentangling superconducting and magnetic orders in NaFe$_{1-x}$Ni$_{x}$As using muon spin rotation}
	
	\author{Sky C. Cheung}
	\thanks{These authors contributed equally to this work.}
	\affiliation{Department of Physics, Columbia University, New York, NY 10027 USA}
	\author{Zurab Guguchia}
	\thanks{These authors contributed equally to this work.}
	\affiliation{Department of Physics, Columbia University, New York, NY 10027 USA}
	\author{Benjamin A. Frandsen}
	\affiliation{Department of Physics, Columbia University, New York, NY 10027 USA}
	\author{Zizhou Gong}
	\affiliation{Department of Physics, Columbia University, New York, NY 10027 USA}
	\author{Kohtaro Yamakawa}
	\affiliation{Department of Physics, Columbia University, New York, NY 10027 USA}
	
	\author{Dalson E. Almeida}
	\affiliation{UEMG Unidade Passos, Av. Juca Stockler, 1130, CEP 37900-106 Passos, MG, Brazil}
	\author{Ifeanyi J. Onuorah}
	\affiliation{Department of Mathematical, Physical and Computer Sciences, Parco delle Scienze 7A, I-43124 Parma, Italy}
	\author{Pietro Bonf\'{a}}
	\affiliation{CINECA, Casalecchio di Reno 6/3 40033 Bologna, Italy}
	\author{Eduardo Miranda}
	\affiliation{Instituto de F\'isica Gleb Wataghin, Unicamp, Rua S\'ergio Buarque de Holanda, 777, CEP
		13083-859 Campinas, SP, Brazil}
	
	\author{Weiyi Wang}
	\affiliation{Department of Physics and Astronomy, Rice University, Houston, Texas 77005, USA}
	\author{David W. Tam}
	\affiliation{Department of Physics and Astronomy, Rice University, Houston, Texas 77005, USA}
	\author{Yu Song}
	\affiliation{Department of Physics and Astronomy, Rice University, Houston, Texas 77005, USA}
	\author{Chongde Cao}
	\affiliation{Department of Physics and Astronomy, Rice University, Houston, Texas 77005, USA}
	\affiliation{Department of Applied Physics, Northwestern Polytechnical University, Xian 710072, China}
	
	\author{Yipeng Cai}
	\affiliation{Department of Physics and Astronomy, McMaster University, Hamilton, ON L8S 4M1 Canada}
	\author{Alannah M. Hallas}
	\affiliation{Department of Physics and Astronomy, McMaster University, Hamilton, ON L8S 4M1 Canada}
	\author{Murray N. Wilson}
	\affiliation{Department of Physics and Astronomy, McMaster University, Hamilton, ON L8S 4M1 Canada}
	\author{Timothy J.S. Munsie}
	\affiliation{Department of Physics and Astronomy, McMaster University, Hamilton, ON L8S 4M1 Canada}
	\author{Graeme Luke}
	\affiliation{Department of Physics and Astronomy, McMaster University, Hamilton, ON L8S 4M1 Canada}
	\author{Bijuan Chen}
	\affiliation{Beijing National Laboratory for Condensed Matter Physics; Institute of Physics,Chinese Academy of Sciences; School of Physics, University of Chinese Academy of Sciences, Beijing 100190, China}
	\author{Guangyang Dai}
	\affiliation{Beijing National Laboratory for Condensed Matter Physics; Institute of Physics,Chinese Academy of Sciences; School of Physics, University of Chinese Academy of Sciences, Beijing 100190, China}
	\author{Changqing Jin}
	\affiliation{Beijing National Laboratory for Condensed Matter Physics; Institute of Physics,Chinese Academy of Sciences; School of Physics, University of Chinese Academy of Sciences, Beijing 100190, China}
	
	
	\author{Shengli Guo}
	\affiliation{Department of Physics, Zhejiang University, Hangzhou 310027, China}
	\author{Fanlong Ning}
	\affiliation{Department of Physics, Zhejiang University, Hangzhou 310027, China}
	
	\author{Rafael M. Fernandes}
	\affiliation{School of Physics and Astronomy, University of Minnesota, Minneapolis, Minnesota 55455, USA}
	\author{Roberto De Renzi}
	\affiliation{Department of Mathematical, Physical and Computer Sciences, Parco delle Scienze 7A, I-43124 Parma, Italy}
	\author{Pengcheng Dai}
	\affiliation{Department of Physics and Astronomy, Rice University, Houston, Texas 77005, USA}
	\author{Yasutomo J. Uemura}
	\email{tomo@lorentz.phys.columbia.edu}
	\affiliation{Department of Physics, Columbia University, New York, NY 10027 USA}
	
	\date{\today}
	
	\begin{abstract}
		Muon spin rotation and relaxation studies have been performed on a ``111'' family of iron-based
		superconductors, \NFNA{}, using single crystalline samples with Ni concentrations $x = 0$, 0.4, 0.6,
		1.0, 1.3, and 1.5\%. Static magnetic order was characterized by obtaining the temperature and doping
		dependences of the local ordered magnetic moment size and the volume fraction of the magnetically
		ordered regions. For $x = 0$ and 0.4\%, a transition to a nearly-homogeneous long range magnetically
		ordered state is observed, while for $x \gtrsim 0.4\%$ magnetic order becomes more disordered and is
		completely suppressed for $x = 1.5$\%. The magnetic volume fraction continuously decreases with
		increasing $x$. Development of superconductivity in the full volume is inferred from Meissner
		shielding results for $x \gtrsim 0.4\%$. The combination of magnetic and superconducting volumes
		implies that a spatially-overlapping coexistence of magnetism and superconductivity spans a large
		region of the $T$-$x$ phase diagram for \NFNA{}. A strong reduction of both the ordered moment size
		and the volume fraction is observed below the superconducting \TC{} for $x = 0.6$, 1.0, and 1.3\%,
		in contrast to other iron pnictides in which one of these two parameters exhibits a reduction below
		\TC{}, but not both. The suppression of magnetic order is further enhanced with increased Ni doping,
		leading to a reentrant non-magnetic state below \TC{} for $x = 1.3$\%. The reentrant behavior
		indicates an interplay between antiferromagnetism and superconductivity involving competition for
		the same electrons. These observations are consistent with the sign-changing $s^{\pm}$
		superconducting state, which is expected to appear on the verge of  microscopic coexistence and
		phase separation with magnetism. We also present a universal linear relationship between the local
		ordered moment size and the antiferromagnetic ordering temperature \TN{} across a variety of
		iron-based superconductors. We argue that this linear relationship is consistent with an
		itinerant-electron approach, in which Fermi surface nesting drives antiferromagnetic ordering. In
		studies of superconducting properties, we find that the $T=0$ limit of superfluid density follows the linear trend observed in underdoped
		cuprates when plotted against \TC{}. This paper also includes a detailed theoretical prediction of
		the muon stopping sites and provides comparisons with experimental results.
	\end{abstract}
	
	
	\maketitle
	
	\section{Introduction}
	\label{sec:Introduction}
	
	Iron-based high temperature superconductors (Fe-HTS) are materials exhibiting unconventional
	superconductivity that arise from parent compounds with static antiferromagnetic (AFM)
	order~\cite{Stewart_FeSC_RMP_2011, Scalapino_FeSC_RMP_2012, Dai_FeSC_RMP_2015}. One of the grand
	challenges in understanding the behavior of these systems is determining the physical mechanism
	responsible for superconductivity. Essential information on the nature of superconductivity in
	strongly correlated electron systems can be deduced by investigating their phase diagrams as well as
	the superconducting (SC) gap structure.
	
	In the parent compound of many Fe-HTS, a spin density wave forms with spins ordered antiparallel to
	each other along one Fe-Fe axis and parallel to each other along the orthogonal Fe-Fe bond
	direction~\cite{RosenthalAndrade_NaFeAs_NPhys_2014, Stewart_FeSC_RMP_2011, Scalapino_FeSC_RMP_2012}.
	Carrier doping, isovalent chemical substitution, or application of pressure to the parent system
	suppresses magnetic order and begets a SC dome~\cite{Uemura_SC_NMat_2009}. In
	addition to magnetism and superconductivity, Fe-HTS exhibit a tetragonal-to-orthorhombic structural
	distortion at a temperature \TS{} that precedes or occurs concurrently with the magnetic phase
	transition at temperature \TN{}~\cite{RosenthalAndrade_NaFeAs_NPhys_2014, Stewart_FeSC_RMP_2011,
		Dai_FeSC_RMP_2015, YiLu_NematicFeSC_PNAS_2011, FernandesChubukov_NematicFeSC_NPhys_2014}. The
	prominent in-plane anisotropy in resistivity along orthogonal axes in the paramagnetic (PM)
	orthorhombic state is associated with an electronic nematic order parameter that triggers the
	orthorhombic distortion of the crystal~\cite{ChuAnalytis_ResistivityFeSC_Science_2010,
		NandiKim_Nematic_PRL_2010}. The aforementioned orders are found in close proximity with each other.
	AFM and SC orders homogeneously coexist in several Fe-HTS, such as in
	\BFCoA{}~\cite{GoltzZinth_BaKFeCoAs2_MagSC_PRB_2014, TamSong_BaFeCoAs2_PRB_2017},
	\BFNA{}~\cite{Arguello_BaFeNiAs2_PhD_2015} and \BKFA{}~\cite{Wiesenmayer_BaKFe2As2_MagSC_PRL_2011}.
	In these systems, the ordered magnetic moment size and nematic order parameter smoothly decrease as
	the temperature is lowered below \TC{}, corroborating the fact that superconductivity and magnetic
	long range order compete for the same electrons \cite{FernandesPratt_PairingFeSC_PRB_2010}. However,
	other studies~\cite{Goko_PRB_2009, Luetkens_NatureMat_2009} have detected the mutual exclusion of
	these two order parameters, i.e. they exhibit macroscopic phase separation in different parts of the
	sample. Characterizing common features of the complex interplay among magnetic, nematic, and SC
	orders in various Fe-HTS is essential for elucidating the microscopic pairing mechanism in Fe-HTS
	and other unconventional superconductors.
	
	One of the major experimental challenges in teasing apart AFM and SC orders is that
	individual experimental probes have limited ranges of sensitivity to magnetism and/or
	superconductivity. For instance, neutron scattering and magnetic susceptibility measurements can
	only reveal volume-integrated information about the magnetic and SC features of the specimens. At
	present, no individual experimental probe can unambiguously address the issue of whether the
	coexistence of AFM and SC orders directly overlap in real space or if the specimen undergoes
	macroscopic phase separation between two phases. In an attempt to clear this experimental hurdle, a
	detailed multiple-probe investigation was recently conducted on \BFNA{}, involving Muon Spin
	Rotation (\MuSR{}), Scanning Tunneling Microscopy (STM), M\"{o}ssbauer spectroscopy, neutron scattering,
	and specific heat measurements~\cite{Arguello_BaFeNiAs2_PhD_2015}. The results from this study offer
	convincing evidence that the AFM and SC phases in \BFNA{} almost completely overlap in real space
	and both phases compete for the same electrons. The question of whether a similar style of phase
	coexistence exists in other families of Fe-HTS remains unclear. In this work, we present a detailed
	\MuSR{} investigation in context with recent susceptibility and neutron scattering measurements to gain a
	deeper understanding of the interplay between AFM and SC orders in \NFNA{}, a member of the ``111''
	family of Fe-HTS.
	
	Recent neutron scattering experiments on \NFNA{} show that the neutron magnetic order parameter is
	diminished below \TC{}~\cite{WangSong_Neutron_2017}, which was interpreted as the reduction of the
	magnetic moment below \TC{}. Using the volume sensitive \MuSR{} technique, we demonstrate for the
	first time in single crystalline samples of \NFNA{}, with $x = 0.6$, 1.0, 1.3, and 1.5\%, that the
	reduction of magnetic intensity is due to a strong reduction of \emph{both} the ordered moment and
	the magnetic volume fraction below \TC{}. The debilitating effect of superconductivity on magnetism
	intensifies as the doping level $x$ increases, leading to a reentrant non-AFM state below \TC{} for
	$x = 1.3\%$. These results suggest an interesting scenario, in which
	the degree of competition between AFM and SC may be itself intrinsically inhomogeneous, varying as a
	function of position in the sample. Moreover, we establish a robust linear dependence between the
	ordered moment and the AFM ordering temperature \TN{} for various Fe-HTS, which is consistent with a
	model of itinerant magnetism in Fe-HTS.
	
	This work is organized as follows: Section~\ref{sec:ExperimentalMethods} describes the preparation
	and handling of the specimens, dc-susceptibility characterization, and the \MuSR{} experimental
	setup. Experimental zero-field \MuSR{} results are shown in Section~\ref{sec:ZFMuSR} and compared
	with neutron scattering results in Section~\ref{sec:NeutronComparison}. A discussion of these
	results is presented in Section~\ref{sec:MagnetismDiscussion}.
	Section~\ref{sec:MuonStoppingSiteFieldSimulation} introduces a muon stopping site simulation
	performed to account for the multiple internal magnetic fields observed in the zero-field \MuSR{}
	spectra. Knowledge of the muon site locations enables the ordered moment size to be determined from
	the observed precession frequency. Section~\ref{sec:Freq_TN_Linear} describes a universal linear
	relation between the ordered moment size and \TN{}. A theoretical discussion of this result
	using a model of antiferromagnetism in Fe-HTS parent compounds based on an itinerant electron
	picture is also presented in this section. Section~\ref{sec:TFMuSR} shows \MuSR{} measurements under
	a transverse external field on superconducting \NFNA{} and demonstrates that a linear relationship
	between the superfluid density and \TC{} is observed in \NFNA{} and other high-\TC{} cuprate
	superconductors. These results are summarized in the concluding Section~\ref{sec:Conclusion}.
	Appendix~\ref{sec:Computational} describes detailed methods and results of the internal field
	simulation. A calculation for the universal scaling of the ordered moment size and ordering
	temperature based on a two-band model is presented in Appendix~\ref{sec:Theory_FreqTN}. Finally,
	Appendix~\ref{sec:SCGap} provides a derivation of the superconducting gap symmetry from the
	temperature dependence of the penetration depth.
	
	\begin{figure*}[ht!]
		\centering
		\includegraphics[width=1\textwidth]{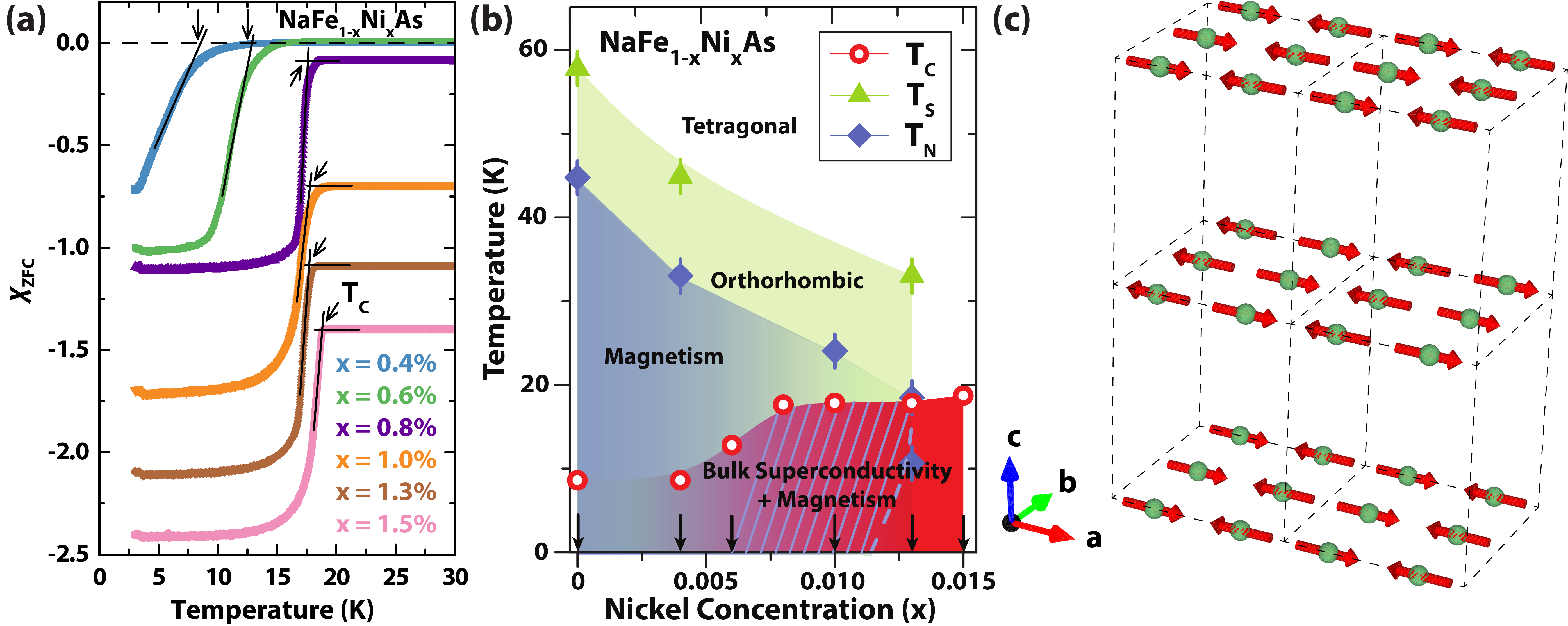}
		\linespread{1}\selectfont{}
		\caption{
			\label{fig:Characterization}
			Magnetic characterization of \NFNA{}.
			(a) Temperature-dependent DC susceptibility measurements in a magnetic field of 5 Oe applied in
			the ab plane. Full SC volume is obtained for $x \ge 0.4$\%. Susceptibility spectra for $x \ge
			0.8\%$ are vertically offset for visual clarity. Black solid lines show how \TC{} (indicated by
			black arrows) was determined for each doping.
			(b) Phase diagram of \NFNA{} illustrating temperature and doping dependences of various orders,
			with structural and magnetic transitions obtained from Ref.~\onlinecite{WangSong_Neutron_2017} and
			displayed as fully-colored symbols. Superconducting transition temperatures \TC{} were determined
			from magnetic susceptibility measurements shown in (a). Black arrows indicate the doping concentrations
			measured by \MuSR{} in our present investigation.
			(c) Collinear AFM spin structure of the undoped compound
			\NFA{}~\cite{LiCruz_NaFeAs_Neutron_PRB_2009, VESTA3_2011}. Only Fe atoms (green) are shown for
			visual clarity. Dashed lines indicate the boundaries of a single unit cell of the crystal structure.
		}
	\end{figure*}
	
	\section{Experimental Methods}
	\label{sec:ExperimentalMethods}
	
	Pristine single-crystal specimens of \NFNA{} with $x = 0$, 0.4, 0.6, 0.8, 1.0, 1.3 and 1.5\% were
	prepared using the self-flux technique in accordance with
	Ref.~\onlinecite{TanatarSpyrison_NaFeAs_PRB_2012}, with each crystallite measuring about $1 \times 1
	\times 0.2$ mm$^3$. Zero-field cooling DC susceptibility measurements were performed on these
	samples in an applied field of 5 Oe in the basal $a$-$b$ plane down to 3 K and the results are shown in
	Figure~\ref{fig:Characterization}(a). These measurements indicate that \NFNA{} exhibits bulk
	superconductivity with full SC shielding fraction for the range $x = 0.6 \sim 1.5\%$, with a maximum
	\TC{} $\approx 17$ K achieved for $x = 1.5\%$. A phase diagram summarizing the structural, magnetic,
	and SC transitions is shown in Figure~\ref{fig:Characterization}(b), which is reminiscent of the
	electronic phase diagrams of \NFCoA{}~\cite{ParkerSmith_MuSR_PRL_2010, WangLuo_NaFeCoAs_PRB_2012}
	and \NFCA{}~\cite{WangLin_NaFeCuAs_PRB_2013}. For clarity, collinear AFM spin structure of the undoped compound NaFeAs
	is also shown in Fig. 1(c).
	
	Since \NFNA{} is highly air and moisture sensitive, the crystallites were tightly encased in packets
	of Kapton film inside an Ar-filled glovebox. Each crystallite was aligned so that the
	crystallographic $c$-axis was oriented normal to the film packet, without any preferred alignment of
	the basal $ab$ plane. For each doping concentration, packets containing a few large crystal
	specimens were mounted on an ultra-low background sample holder using aluminum tape.
	
	In a \MuSR{} experiment, positive muons implanted into a specimen serve as extremely sensitive local
	probes to simultaneously measure small internal magnetic fields and ordered magnetic volume
	fractions. Therefore, we can ascertain the temperature and doping evolution of the magnetic volume
	fraction and ordered moment separately, unlike reciprocal-space techniques such as neutron
	scattering. Time differential \MuSR{} measurements were performed using the Los Alamos Meson Physics
	Facility (LAMPF) spectrometer with a helium gas-flow cryostat at the M20 surface muon beamline (500
	MeV) of TRIUMF in Vancouver, Canada and using the General Purpose Surface-Muon Instrument (GPS) with
	a standard low-background veto setup at the ${\pi}$M3 beam line of the Paul Scherrer Institute in
	Villigen, Switzerland. A continuous beam of 100\% spin polarized muons was implanted into the sample
	and the time dependence of the ensemble muon polarization was collected at temperatures between 2K
	and 70K. The muon beam momentum was parallel to the crystal $c$ axis. By applying magnetic fields to the
	muon beam before the sample, the ensemble muon spin prior to implantation can be oriented parallel
	or perpendicular to the beam direction. See Refs.~\onlinecite{MuSR:YaouancDalmas, MuSR:Schenck,
		MuSR:MuonScience} for further details on the \MuSR{} experimental technique. The \MuSR{} spectra
	were analyzed in the time domain using least-squares optimization routines from the \texttt{musrfit}
	software suite~\cite{Suter_MuSRFit_2012}.
	
	\section{Magnetism in $\textrm{NaFe}_{1-x}\textrm{Ni}_{x}\textrm{As}$}
	\label{sec:Magnetism}

	\subsection{Zero Field \MuSR{} Results}
	\label{sec:ZFMuSR}
	
	\begin{figure*}[ht!]
		\centering
		\includegraphics[width=0.99\textwidth]{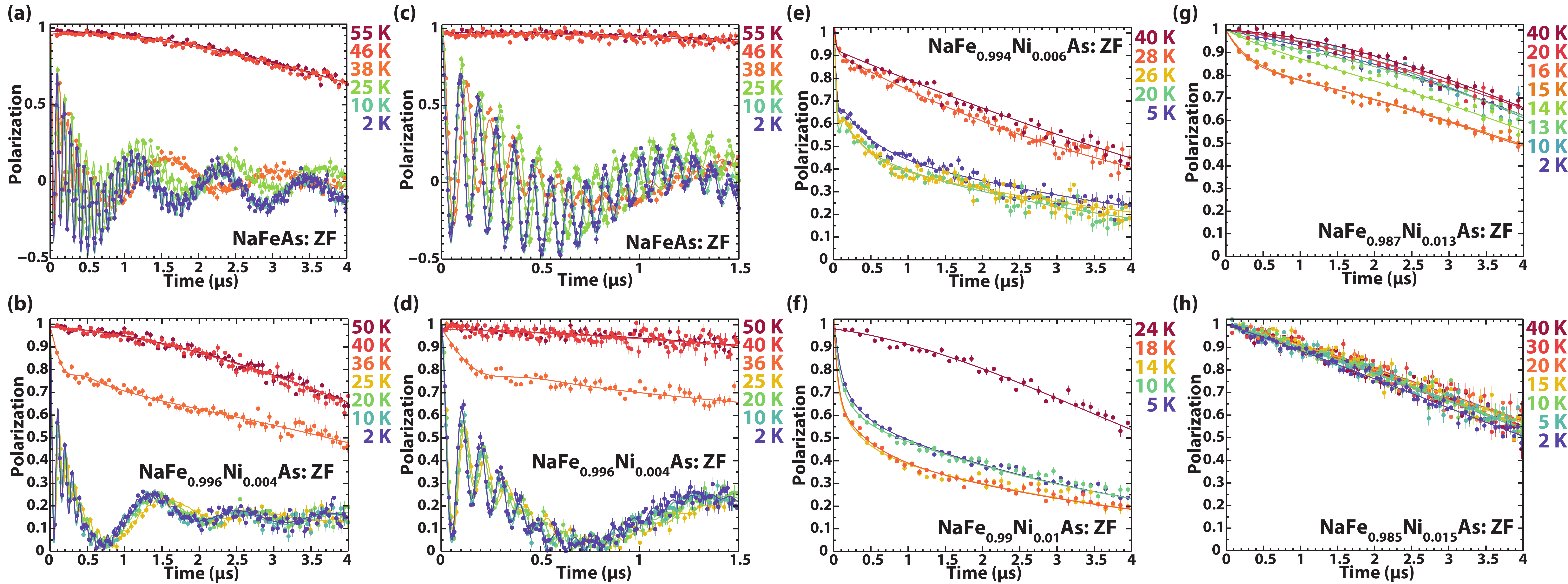}
		\linespread{1}\selectfont{}
		\caption{
			\label{fig:ZF_Spectra}
			ZF-\MuSR{} spectra on \NFNA{}. 
			(a)-(b) Muon spin polarization in zero field for \NFNA{} for $x = 0$ and 0.4\%, respectively.
			(c)-(d) Zoomed-in view of the first 1.5 microseconds of the spectra shown in (a) and (b). (e)-(h)
			Time spectra for the $x = 0.6$, 1.0, 1.3, and 1.5\% compound in zero field, respectively. Solid
			lines in all panels are fits of the data to the ZF-\MuSR{} model in~\eqref{eq:ZFPolarizationFit}
			for each temperature. Additional details on the \MuSR{} time spectra and experimental geometry can
			be found in Refs.~\onlinecite{MuSR:YaouancDalmas, MuSR:Schenck, MuSR:MuonScience}.
		}
	\end{figure*}
	
	The observed \MuSR{} time spectra (muon ensemble polarization) of $x = 0$, 0.4, 0.6, 1.0, 1.3 and
	1.5\% in zero applied field (ZF-\MuSR{}) are shown in Figure~\ref{fig:ZF_Spectra}. In these
	measurements, the initial muon spin polarization is in the $a$-$b$ plane of the crystals, and
	the time spectra were obtained using up and down positron counters. At high temperatures, only a
	very faint depolarization of the \MuSR{} signal is observed. This weak relaxation mostly originates
	from the interaction of the muon spin with randomly oriented nuclear magnetic moments. Upon cooling,
	the relaxation of the \MuSR{} signal increases due to the proliferation of Fe-spin correlations.
	
	For $x = 0$ and 0.4\% samples, three distinct precession frequencies occur in the \MuSR{} spectra,
	which emanate from three magnetically inequivalent muon stopping sites in \NFNA{}, in agreement with
	our stopping site calculations presented in Section~\ref{sec:MuonStoppingSiteFieldSimulation}. No
	coherent oscillations are present in the $x \gtrsim 0.6\%$ spectra shown in
	Figure~\ref{fig:ZF_Spectra}(e)-(f), even at the lowest measured temperature, as only a rapidly
	relaxing signal is observed. The fast depolarization of the \MuSR{} signal (without oscillations)
	arises from a broad distribution of static internal magnetic fields, which has been confirmed using
	longitudinal field (LF)-\MuSR{} experiments. These measurements reveal that the muon spin relaxation
	is substantially suppressed at modest longitudinal external fields between 25 and 50 mT (of the
	order of internal quasistatic fields), suggesting an inhomogeneous magnetic state in the samples
	with $x = 0.6$, 1.0 and 1.3\%. The ZF-\MuSR{} time spectra for the $x = 1.3\%$ compound shown in
	Figure~\ref{fig:ZF_Spectra}(g) demonstrate magnetic ordering between 14 K and 17 K. Below 14 K,
	magnetic order vanishes and the specimen only exhibits bulk superconductivity. Interestingly, a
	similar re-entrance to a non-magnetic state was observed in BaFe$_{2-x}$Co$_x$As$_2$ by neutron
	diffraction~\cite{FernandesPratt_PairingFeSC_PRB_2010}. In the following, we present how the
	magnetic properties of \NFNA{} evolve with temperature and doping.
	
	All of the ZF-\MuSR{} spectra were fit to the following phenomenological model:
	
	
	\begin{widetext}
		\begin{equation}
		\label{eq:ZFPolarizationFit}
		P_{\textrm{ZF}}(t) = F \left( \sum_{j = 1}^{3}  \left( f_j \cos(2\pi\nu_j t + \phi) e^{-\lambda_j t} \right) +
		f_L e^{-\lambda_{L} t} \right) + (1-F) \left(\frac{1}{3} + \frac{2}{3}\left( 1 - \lambda t - (\sigma t)^2  \right) e^{-\lambda t-\frac{1}{2}(\sigma t)^2} \right)
		\end{equation}
	\end{widetext}
	The model in~\eqref{eq:ZFPolarizationFit} consists of an anisotropic magnetic contribution
	characterized by an oscillating ``transverse'' component and a slowly relaxing ``longitudinal''
	component.  The longitudinal component arises due to the parallel orientation of the muon spin
	polarization and the local magnetic field. In polycrystalline samples with randomly oriented fields
	this results in a so-called ``one-third tail'' with $f_L = \frac{1}{3}$. For single crystals, $f_L$
	varies between zero and unity as the orientation between field and polarization changes from being
	parallel to perpendicular.  In addition to the magnetically ordered contribution, there is a PM
	signal component characterized by the densely distributed network of nuclear dipolar moments
	$\sigma$ and dilute electronic moments with random orientations
	$\lambda$~\cite{GuguchiaKhasanov_PRB_2016}. The temperature-dependent magnetic ordering fraction $0
	\le F \le 1$ governs the trade-off between magnetically-ordered and PM behaviors. 
	
	\begin{figure}[ht!]
		\centering
		\includegraphics[width=0.98\columnwidth]{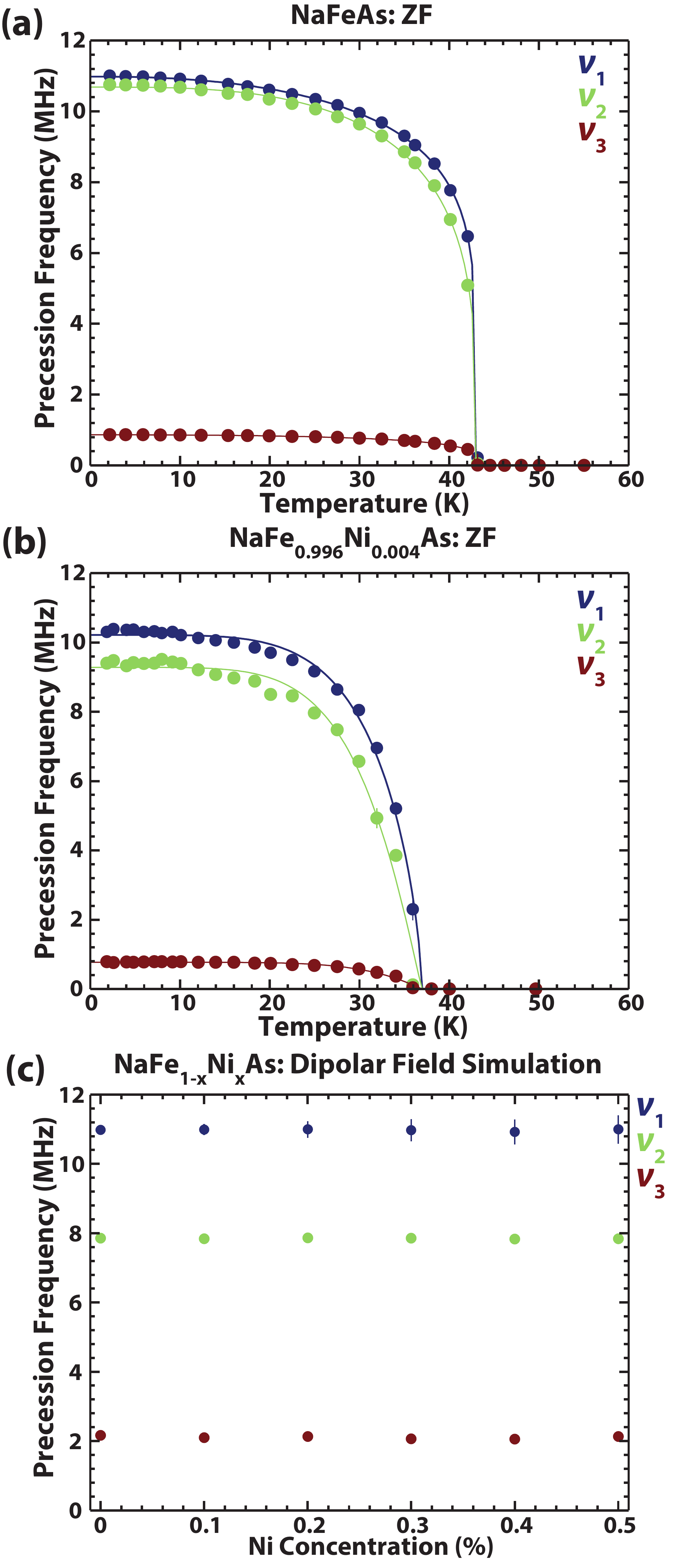}
		\linespread{1}\selectfont{}
		\caption{ 
			\label{fig:ZF_Frequency}
			Muon precession frequencies in \NFNA{}. 
			(a)-(b) Precession frequencies $\nu_j$ from the model used in~\eqref{eq:ZFPolarizationFit} on the
			$x = 0$ and 0.4\% compounds, respectively. Solid lines are power-law fits to the data. (c)
			Simulation results from dipolar field calculations on lowly-doped \NFNA{} using the three muon
			stopping sites obtained from DFT calculations.
		}
	\end{figure}
	
	Shown in Figure~\ref{fig:ZF_Frequency}(a)-(b) are the temperature dependences of the precession
	frequencies observed in the $x = 0$ and 0.4\% samples. For the undoped and $x = 0.4\%$ systems,
	there are three distinct frequencies that share the same relationship  $\nu(T) = \nu(0) ( 1 - (
	\frac{T}{T_N} )^{a} )^{b}$, which are indicated by solid lines. In the parent system, a sharp
	step-like increase of $\nu(T)$ is observed below \TN{} $ \approx 42 $ K, which may be a signature of
	a first-order phase transition, although further experiments are needed to establish the character
	of the transition. This feature is absent in the $x = 0.4\%$ sample, which could be due to disorder
	effects introduced by Ni impurities~\cite{Goko_NPJQM_2017}. Similar ZF-\MuSR{} experiments were also
	performed by using positron counters located in the forward and backward directions with respect to
	the muon beam direction. Interestingly, the two fast frequencies are absent in the non-spin-rotated
	spectra for $x = 0$ and $0.4\%$. If we associate each frequency to a different muon stopping site,
	these results suggest that the magnetic field directions at the high-field stopping sites are
	oriented along $c$ axis of the crystal. This feature is consistent with dipolar field simulations on
	muon stopping sites presented in Section~\ref{sec:MuonStoppingSiteFieldSimulation}.
	
	We define the static magnetic order parameter $\mathcal{M} \equiv \sqrt{(2\pi\nu)^2 +
		\lambda_{\textrm{T}}^2}$ to track the temperature and doping dependence of magnetism, where $\nu$ is
	the maximum precession frequency and $\lambda_{\textrm{T}}$ is the relaxation rate corresponding to
	$\nu$. As defined, $\mathcal{M}$ takes into account both homogeneous (well-defined precession
	frequency $\nu$) and inhomogeneous contributions (rapid early-time relaxation
	$\lambda_{\textrm{T}}$) to the signal. Therefore, the magnetic transition temperature \TN{}
	corresponds to the onset of $\mathcal{M}$.
	
	
	\begin{figure}[ht!]
		\centering
		\includegraphics[width=0.99\columnwidth]{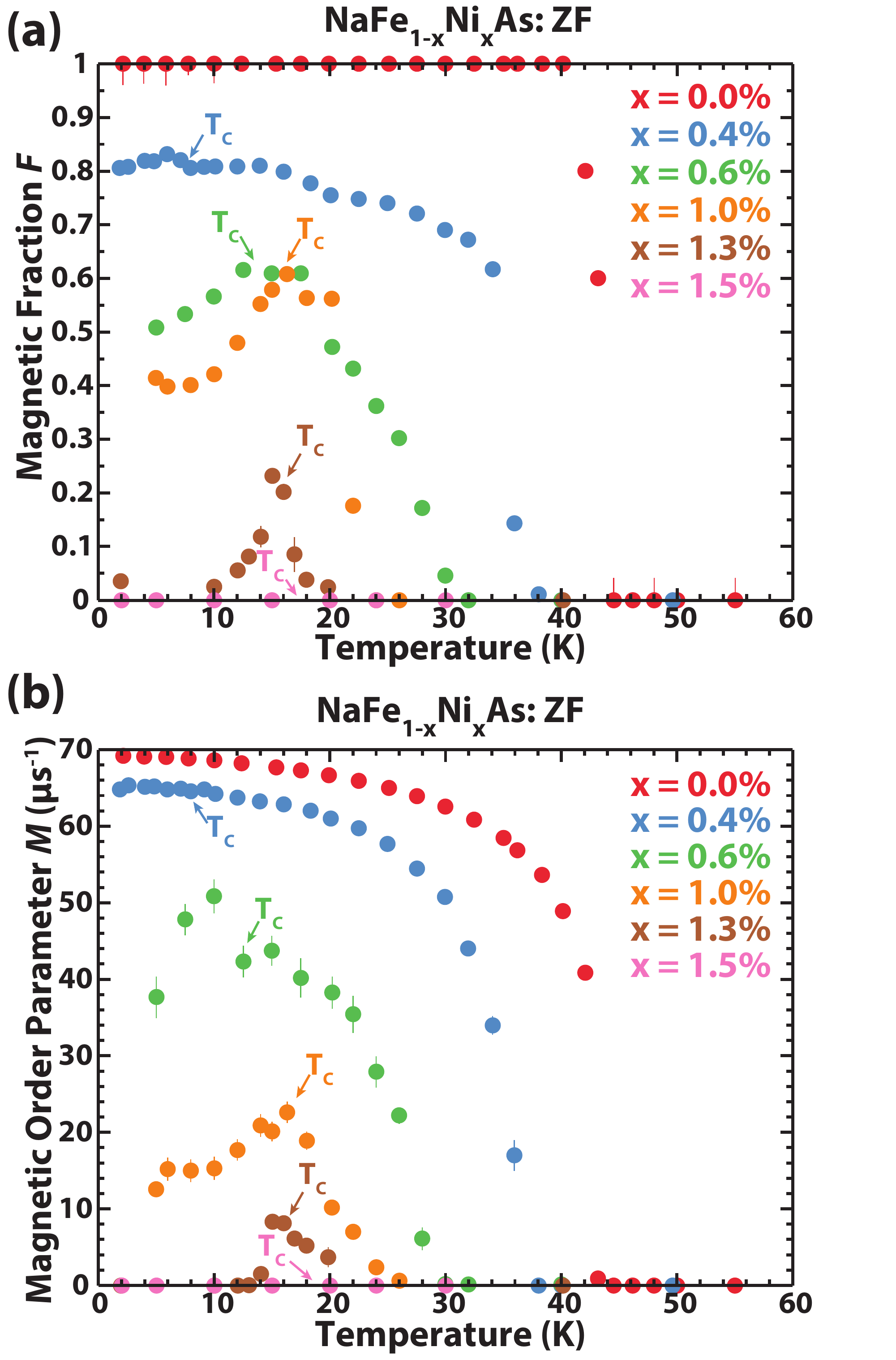}
		\linespread{1}\selectfont{}
		\caption{ 
			\label{fig:ZF_FitParameters}
			Summary of ZF-\MuSR{} fit results on \NFNA{}. 
			(a) Magnetically ordered volume fraction $(F)$ as a function of temperature and Ni concentration.
			(b) Static magnetic order parameter $(\mathcal{M})$ as a function of temperature and Ni
			concentration. Arrows in the figures denote \TC{} for the various samples.
		}
	\end{figure}
	
\begin{figure}[ht!]
		\centering
		\includegraphics[width=0.95\columnwidth]{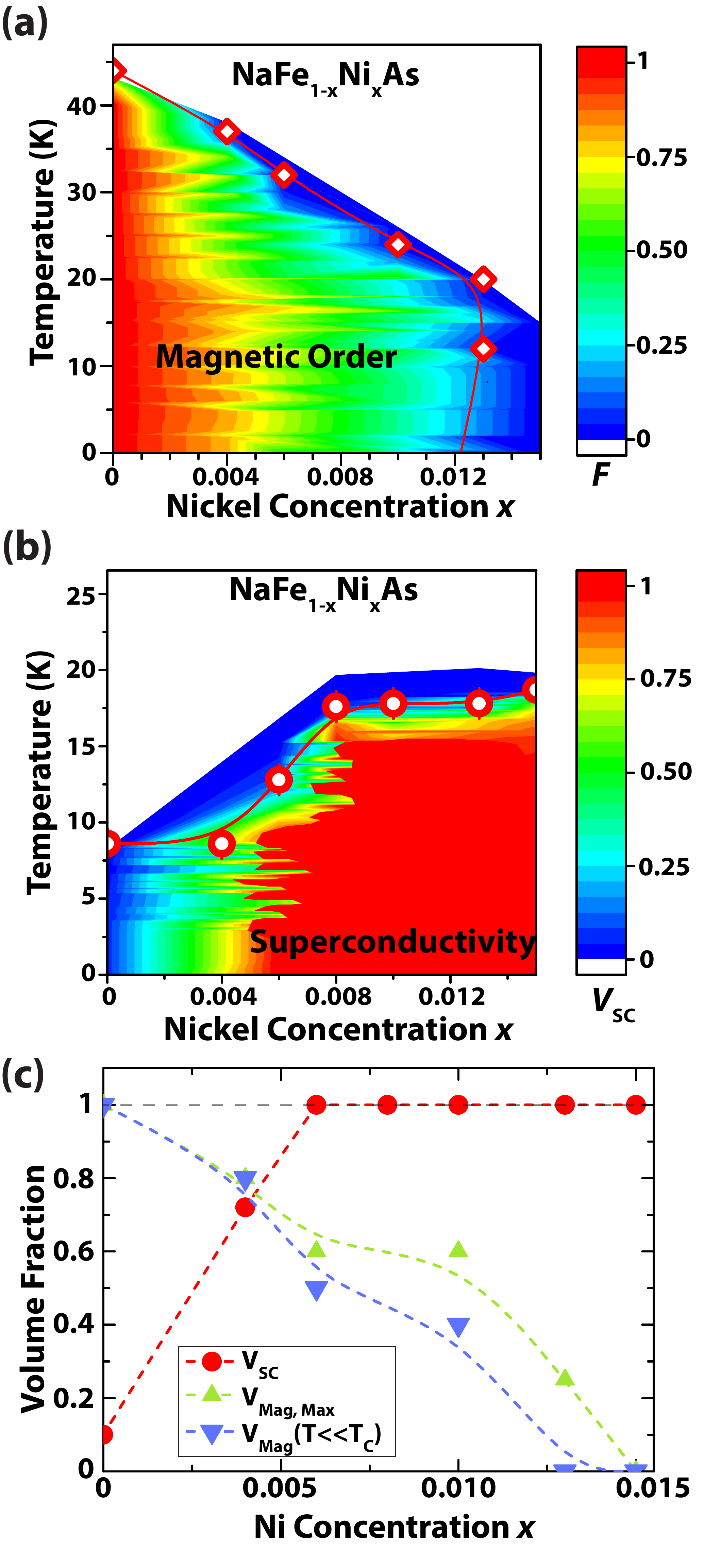}
		\linespread{1}\selectfont{}
		\caption{
			\label{fig:MuSR_Summary_PhaseDiagram}
			Magnetic and SC volume fractions in \NFNA{}.
			(a) Doping and temperature evolution of the magnetic fraction $V_{\textrm{Mag}} = F$ from ZF-\MuSR{}. Red diamonds indicate \TN{} and the red curve is a guide to the eye. Observe the bend in the curve near $x = 0.013$ indicating a reentrant non-AFM phase.
			(b) Doping and temperature evolution of the SC volume fraction from magnetic susceptibility
			measurements presented in Figure~\ref{fig:Characterization}(a). Red circles indicate \TC{} and the red curve is a guide to the eye.
			(c) Summary of magnetic and SC volume fractions. $V_{SC}$ is the SC volume fraction,
			$V_{\textrm{Mag, Max}}$ is the maximum value of $F$ for each doping, and $V_{\textrm{Mag}}(T \ll
			T_{\textrm{C}})$ is $F$ at temperatures much less than \TC{}.
		}
	\end{figure}

	The temperature and doping evolution of the magnetic fraction $F$ and magnetic order parameter
	$\mathcal{M}$ are shown in Figure~\ref{fig:ZF_FitParameters}. The relative decrease in $\mathcal{M}$
	below \TC{} is more pronounced with increased doping, as seen in
	Figure~\ref{fig:ZF_FitParameters}(b). Indeed, the $x = 1.3\%$ sample exhibits reentrant behavior in
	which the low-temperature state becomes non-AFM below \TC{} within experimental uncertainty. The
	temperature evolution of the magnetically ordered fraction $F$ is shown in
	Figure~\ref{fig:ZF_FitParameters}(a). $F$ shows a sharp increase below \TN{} while the onset of
	SC causes $F$ to decrease when cooled below \TC{}. With higher doping, a stronger reduction of $F$
	is observed below \TC{}. For the $x = 1.3\%$ system, magnetic order is completely destroyed and the
	system loses long-range AFM order below 14 K. A summary of the and temperature doping dependences of the magnetic
	and SC volume fractions is presented in Figure~\ref{fig:MuSR_Summary_PhaseDiagram}.
	
	\subsection{Comparison with Elastic Neutron Scattering}
	\label{sec:NeutronComparison}
	
	\begin{figure}[ht!]
		\centering
		\includegraphics[width=0.99\columnwidth]{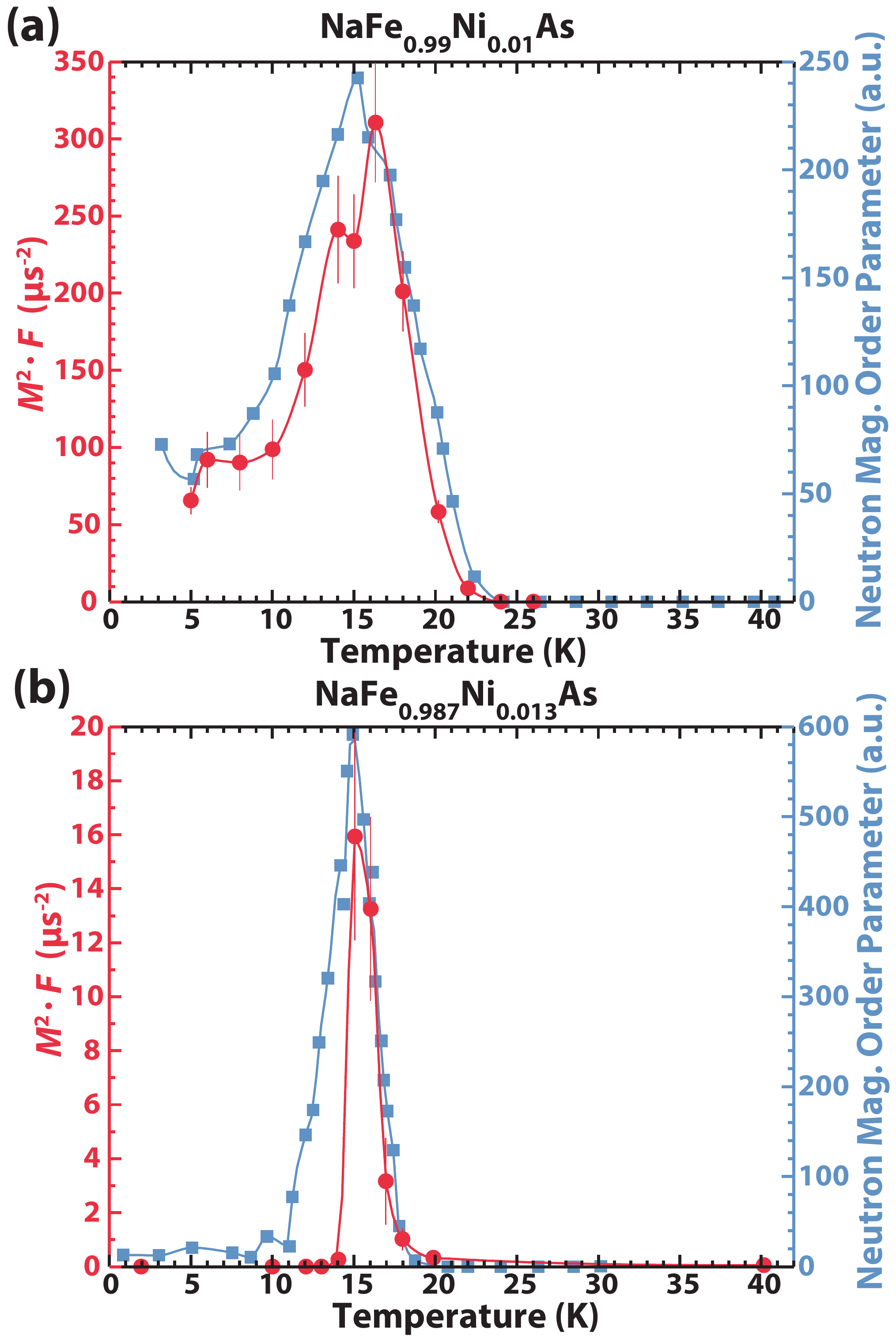}
		\linespread{1}\selectfont{}
		\caption{ 
			\label{fig:MuSR_Neutron_Comparison}
			Comparison of \MuSR{} and elastic neutron scattering measurements~\cite{WangSong_Neutron_2017} for
			the (a) $x = 1.0\%$ and (b) $x = 1.3\%$ compounds. The magnetic strength $\mathcal{M}^2 F$ from
			\MuSR{} measurements is plotted in red and the Bragg peak intensity from neutron scattering is shown
			in blue.
		}
	\end{figure}
	
	As mentioned in Section~\ref{sec:Introduction}, elastic neutron scattering experiments show that the
	neutron magnetic order parameter is diminished below \TC{}~\cite{WangSong_Neutron_2017} in \NFNA{},
	which was interpreted as the reduction of the magnetic moment below \TC{}. A comparison between the
	neutron magnetic order parameter and the magnetic strength $\mathcal{M}^2 F$ from our \MuSR{}
	studies is shown in Figure~\ref{fig:MuSR_Neutron_Comparison} for the $x = 1.0$ and 1.3\% systems. As
	a volume-integrating probe in reciprocal space, neutron scattering techniques are sensitive to both
	the ordered moment and its volume fraction, but these two contributions cannot be separated from the
	measured scattered intensity. Consequently, the suppression of magnetic order below \TC{} observed
	in neutron diffraction cannot be unambiguously attributed to a reduction of the magnetic moment.
	However, \MuSR{} enables independent measurements of both the volume fraction and the ordered moment
	size, unlike neutron scattering and other bulk probes. From our ZF-\MuSR{} results in
	Figure~\ref{fig:ZF_FitParameters}, we conclude that the suppression of magnetic ordering is due to a
	decrease in both the ordered volume fraction and the moment size.
	
	\subsection{Discussion}
	\label{sec:MagnetismDiscussion}
	
	Our results offer strong evidence that both the ordered moment and fraction are partially or fully
	suppressed below \TC{}. The strong suppression of the magnetism below the onset of superconductivity
	was also observed in the ``122'' and ``11'' families of Fe-HTS:
	\BFCoA{}~\cite{GoltzZinth_BaKFeCoAs2_MagSC_PRB_2014, TamSong_BaFeCoAs2_PRB_2017} (where re-entrance
	of the non-AFM phase was reported~\cite{FernandesPratt_PairingFeSC_PRB_2010}),
	\BFNA{}~\cite{Arguello_BaFeNiAs2_PhD_2015}, \BKFA{}~\cite{Wiesenmayer_BaKFe2As2_MagSC_PRL_2011}, and
	FeSe~\cite{Bendele_PRB_2012}. However, we note that in \BFCoA{} and \BKFA{}, only the ordered moment
	decreases below \TC{}, but the magnetic fraction remains unaffected. On the other hand, both the
	ordered moment and magnetic fraction decrease below \TC{} for FeSe (which becomes magnetic under
	hydrostatic pressure). Results in the present investigation of \NFNA{} are similar to the FeSe case.
	Itinerant AFM and SC orders are generally expected to compete strongly for the same electronic
	states; this behavior can be captured within a simple Ginzburg-Landau free energy for the AFM and SC
	order parameters, $\mathcal{M}$ and $\Delta$, respectively (in the context of Fe-HTS, see for
	instance Refs.~\onlinecite{FernandesPratt_PairingFeSC_PRB_2010, FernandesSchmalian_FreqTN_PRB_2010,
		VorontsovVavilov_FreqTN_PRB_2010, Almeida_PRB_2017}):
	
	\begin{equation*}
	\label{GL} 
	F = \frac{a_m}{2} \mathcal{M}^2 + \frac{u_m}{4} \mathcal{M}^4 + \frac{a_s}{2} \Delta^2 +
	\frac{u_s}{4} \Delta^4 + \frac{\gamma}{2} \mathcal{M}^2 \Delta^2
	\end{equation*}
	
	The degree of competition between these two orders is encoded in the combination of coefficients $g
	= \gamma/\sqrt{u_s u_m}$. If the competition is too strong $(g>1)$, these two orders are
	macroscopically phase separated and do not coexist microscopically. On the other hand, if the
	competition is weak $(g<1)$, they can establish a coexistence phase in which both order parameters
	are simultaneously non-zero at the same position. In a homogeneous system, the first scenario is
	manifested by a reduction of the AFM volume fraction $F$ below \TC{} without a change in the size of
	the magnetic moment. Conversely, the second scenario is manifested by a reduction of the magnetic
	moment below \TC{} without any variation in the volume fraction. Interestingly, we observe both
	signatures in \NFNA{}. Although a detailed theoretical analysis is beyond the scope of this work,
	this suggests that the parameter $g$ itself may be inhomogeneous and change as a function of the
	position in the sample. If Cooper pairs were to form an unconventional sign-changing $s^{\pm}$
	state~\cite{Hirschfeld_review, Chubukov08}, it was argued~\cite{FernandesPratt_PairingFeSC_PRB_2010}
	that the system would be at the verge of phase separation and microscopic coexistence, i.e. $g
	\approx 1$. In this case, local inhomogeneity could locally alter the value of $g$ in a significant
	manner~\cite{Hoyer14}.
	
	\section{Internal Field Simulations at Muon Stopping Sites}
	\label{sec:MuonStoppingSiteFieldSimulation}
	
	To investigate the effect of Ni-dopants on the magnetism in \NFNA{}, we numerically simulate the
	behavior of the muon in the magnetic environment of \NFNA{}. In low-temperature \MuSR{} experiments, the incident muons
	thermalize with the lattice and are implanted at interstitial locations referred to as stopping
	sites. Muon implantation sites in Fe-HTS have been successfully identified using a succession of
	increasingly accurate theoretical calculations. Early studies were based on the analysis of the
	local minima of the unperturbed electrostatic potential within either the simple Thomas Fermi or a
	full Density Functional Theory (DFT) approach. This strategy was specifically followed for the
	``1111''~\cite{deRenziBonfa_DFT_Muon_1111_SST_2012, PrandaBonfa_1111_DFT_Muon_PRB_2013,
		Luetkens_NatureMat_2009, MaeterLuetkins_DFT_Muon_1111_PRB_2009} and the
	``11''~\cite{KhasanovGuguchia_11_DFT_Muon_PRB_2017, Bendele_PRB_2012} classes of Fe-HTS. In
	addition, similar calculations were performed on selected
	``122''~\cite{SugiyamaNozaki_122_DFT_Muon_PRB_2015, Shermadini_Thesis_DFT_Muon_2014} materials and
	other systems~\cite{BonfaSartori_DFT_Muon_2015, AdamSuprayoga_DFT_Muon_2014}. Recently, and
	exclusively for the ``1111'' family of Fe-HTS, the effect of the muon on the lattice was captured within
	a supercell DFT impurity calculation by considering force and energy relaxations of possible muon
	implantation sites~\cite{MollerBonfa_DFT_Muon_2013, BonfadeRenzi_DFT_Muon_JPSoJ_2016}.
	
	The ab-initio search often identifies clusters of sites. This is true also in the simple unperturbed
	potential method, that fails in insulators such as
	fluorides~\cite{BernardiniBonfa_Fluoride_DFT_Muon_PRB_2013}, but yields a correct first
	approximation in the metallic pnictides owing to the electron screening of the muon charge. In this
	case, the clusters are defined as the portion of the unit cell volume enclosed by the isosurface
	corresponding to the muon ground state energy. More accurate stopping site determination would
	require an impurity DFT approach. Under this methodology, clusters of candidate muon sites are
	generally found with smaller \emph{intra}-cluster and larger \emph{inter}-cluster energy barriers.
	
	Since muon localization is a metastable epithermal kinetic process that cannot be described by a
	mere minimum energy criterion, all of these methods uncover clusters of candidate locations that may
	not directly correspond to observed muon sites. In principle, the true muon fate could be simulated
	by robust \emph{ab initio} path integral molecular dynamics
	~\cite{MarxParrinello_PIMD_DFT_Muon_JCP_1996, MiyakeOgitsu_PIMD_DFT_Muon_PRL_1998}. At present, these
	techniques are computationally prohibitive for impurity calculations on complex structures such as
	Fe-HTS. Therefore, the most feasible method for muon site determination in Fe-HTS involves comparing the
	experimental and calculated local field values at candidate sites.
	
	\subsection{Candidate Muon Stopping Sites in ``111'' Systems}
	
	To determine plausible muon implantation sites in the ``111'' family of Fe-HTS, we employed DFT
	methods that account for local crystal deformations and electronic band structure perturbations due
	to the implanted muons. In particular, muon stopping site calculations in \NFA{} and \NFNA{} were
	performed using spin-polarized DFT with plane wave expansions of the Kohn-Sham orbitals at both
	atomic and interstitial sites. The Generalized Gradient Approximation (GGA) was applied for the
	exchange correlation functional within the Perdew-Burke Ernzerhof (PBE)
	formalism~\cite{PerdewWang_GGA_DFT_PRB_1992, PerdewBurke_GGA_DFT_PRL_1996}. Finally, the core
	wavefunction was approximated using the Projector Augmented Wave (PAW)
	method~\cite{Blochl_PAW_DFT_PRB_1994}. The plane wave and charge density cutoffs were chosen to be
	120 and 1080 Ry, respectively. More details on the muon site determination procedure are found in
	Appendix~\ref{sec:InitializationSimulation} and~\ref{sec:DFT}.
	
	This initial search with DFT methods uncovered five plausible muon sites in \NFA{}, which are also
	assumed to be valid for lowly doped \NFNA{}. The five candidate sites were grouped into
	two clusters based on relative calculated energies. Since the muon is treated as a classical
	particle within DFT, corrections due to its light mass can be included by taking into account the
	spread of the muon wavefunction in the Double Born-Oppenheimer (DBO) approximation method
	~\cite{BonfaSartori_DFT_Muon_2015}. As described in Appendix~\ref{sec:DFT}, we invoked the DBO
	approximation to examine the relative stability of the five candidate sites. From our stability checks,
	we concluded that only two of the three muon sites in the low-energy cluster proved to be stable.
	In addition, both muon sites in the high-energy cluster relax into each other, suggesting that the
	muon is likely delocalized between these two sites, which are also in close proximity to each other.
	As a result, we have determined three plausible muon stopping locations (two
	stopping sites and a delocalized high-energy stopping position) in \NFA{}, which are listed in
	Table~\ref{tab:DFTDipolarSummary}.
	
	\subsection{Dipolar Internal Field Simulations on \NFNA{}}
	
	With the muon stopping sites determined, magnetic dipolar field simulations were performed by
	simulating the \NFA{} as an array of localized magnetic dipoles. The two dipolar contributions
	considered in the internal field simulation are localized electronic moments from AFM-ordered
	Fe atoms, and random nuclear dipolar moments from all atoms. Non-magnetic nickel impurities were
	randomly substituted into the Fe sites on the host \NFA{} lattice to generate \NFNA{}. By performing
	a vector sum of the array of (static) electronic and random nuclear dipolar moments, the internal
	field distribution was numerically simulated for all points in the crystal.
	
	To capture the stochastic fluctuations in the random nuclear moment directions and Ni site
	substitutions, the internal field distribution was simulated by performing 10,000 independent trials
	of dipolar sums for each muon site as a function of Ni concentration $x$. Although simulated results
	can be implemented for any $x$, the simplified dipolar field model severely breaks down beyond $x
	\gtrsim 0.4\%$ since the simulation does not consider bulk superconductivity (see
	Figure~\ref{fig:Characterization}). Magnetic disorder induced by the SC state at $x \gtrsim
	0.4\%$ could also explain the disappearance of coherent oscillations in the ZF-\MuSR{} spectra in
	Figure~\ref{fig:ZF_Spectra}. Additional details on the simulation setup for exploring the local
	magnetic environment at the \NFNA{} stopping sites are found in
	Appendix~\ref{sec:InitializationSimulation} and~\ref{sec:DipolarSimulation}.
	
	\begin{table*}[ht!]
		\centering
		\caption{
			Summary of dipolar field simulations in \NFA{} using muon stopping site positions obtained from
			DFT methods. Similar stopping sites are expected for lowly-doped \NFNA{}. Stopping site positions
			are given in fractional coordinates. The highest frequency from the dipolar field simulations is in close agreement with the experimental results assuming a static ordered Fe moment of 0.175(3)\muB{}.
		}
		\label{tab:DFTDipolarSummary}
		\begin{ruledtabular}
			\begin{tabular}{cccccc}
				Site & Site Position~\footnote{Stopping site positions given in fractional coordinates.} & Simulated
				$\nu$ (MHz)~\footnote{Muon precession frequency from dipolar field simulations.} & Experimental $\nu$
				(MHz)~\footnote{Muon precession frequency from \MuSR{} experiments on \NFA{}} & Simulated $\theta$ $(^\circ)$~\footnote{Average acute angle between the simulated field direction and the $c$-axis} & Experimental $\theta$ $(^\circ)$~\footnote{Average acute angle between the local field direction and the $c$-axis. The local field direction was estimated from ZF-\MuSR{} measurements with the muon spins rotated in orthogonal directions.}\\ \hline
				1 & (0.000, 0.875, 0.100) & 10.987(49) & 10.981(27) & 42.1(5) & 0(10)\\ 
				2 & (0.100, 0.750, 0.100) & 7.839(30) & 10.685(57) & 31.1(6) & 0(10)\\ 
				3~\footnote{From stability analysis of calculated muon sites, the third frequency is likely
					attributed to a stopping site delocalized across sites D and E. See Appendix~\ref{sec:DFT} for
					more details. For simplicity, we list here the simulated results calculated for site class E from
					Table~\ref{tab:MuonSiteDFT} in Appendix~\ref{sec:DFT}. } & (0.500, 0.250, 0.600) & 2.090(21) &
				0.864(06) & 0.6(4) & 18(10)\\
			\end{tabular}
		\end{ruledtabular}
	\end{table*}
	
	\subsection{Discussion of Computational Results}
	
	The main results of our computational investigation are summarized in
	Table~\ref{tab:DFTDipolarSummary}. Our stopping site calculations and stability analysis reveal
	three plausible muon stopping sites in \NFA{}. This is consistent with the observation that there
	are three precession frequencies in the ZF-\MuSR{} spectra in lowly-doped \NFNA{}. The calculated
	precession frequencies are listed in Table~\ref{tab:DFTDipolarSummary}, along with the extrapolated
	frequencies from power law fits of the frequencies from \MuSR{} found in
	Figure~\ref{fig:ZF_Frequency}(a). Moreover, our simulations show that the mean local fields at the
	two high-field sites make an acute angle of approximately $36^{\circ}$ with the crystal $c$ axis,
	implying that the strong fields at these sites are preferentially aligned with the $c$ axis. This is
	consistent with our experimental observation that the high frequency oscillations have noticeable
	amplitudes when the initial muon spin is not aligned with the $c$ axis (i.e. in the spin-rotated
	configuration), as shown in Figure~\ref{fig:ZF_Spectra}(a). Differences in the simulated and
	experimentally-obtained angles $\theta$ suggest that the true muon sites are likely a small
	displacement from the ones listed in Table~\ref{tab:DFTDipolarSummary}.
	
	The doping evolution of the simulated precession frequencies are shown in
	Figure~\ref{fig:ZF_Frequency}(c). Comparisons of the simulated and observed frequencies for Site 1
	in our dipolar field simulations enabled us to estimate the ordered moment size of the Fe atoms
	in \NFA{} to be $\mu_{\textrm{Fe}} = 0.175(3) \mu_{\textrm{B}}$. The difference between the simulated
	and experimental frequencies for the second and third sites suggests that quantum correlations (e.g.
	contact hyperfine fields) contribute to the internal field, which are not included in the dipolar
	model. In addition, the presence of Ni dopants can perturb the ordering of Fe moments, which was not
	included in the simulation. Nonetheless, our computational investigation provides a physical
	interpretation of the frequencies observed in the ZF-\MuSR{} spectra and corroborates the model for
	the magnetic ordering in~\eqref{eq:ZFPolarizationFit}.
	
	The ordered moment size in a variety of Fe-HTS has been explored experimentally using \MuSR{},
	neutron scattering, and M\"{o}ssbauer
	measurements~\cite{LumsdenChristianson_Magnetism_FeHTS_JPCM_2010}. The reported variations of the
	ordered Fe moments of the same specimen is a testament to the differences in sensitivity across
	these three probes of the local moment. Table~\ref{tab:OrderedMomentComparison} shows a comparison
	of the ordered moment size of representative systems from the various classes of Fe-HTS. The
	estimate from our present investigation in \NFA{}, $\mu_{\textrm{Fe}} = 0.175(3) \mu_{\textrm{B}}$,
	is consistent with the moment sizes reported from neutron
	scattering~\cite{TanSong_NaFeCoAs_Neutron_PRB_2016} and M\"{o}ssbauer
	spectroscopy~\cite{PresniakovMorozov_111_Mossbauer_JPCM_2013}.
	
	\begin{table}[ht!]
		\centering
		\caption{
			Comparison of the low-temperature Fe ordered magnetic moments in selected Fe-HTS. All magnetic moments are given in units of $\mu_{\textrm{B}}$.
		}
		\begin{ruledtabular}
			\begin{tabular}{cccc}
				Fe-HTS & \MuSR{} & Neutron Scattering & M\"{o}ssbauer~\footnote{Ordered moment extrapolated from measured low-temperature hyperfine field using the scaling relation $15$ T$/\mu_{\textrm{B}}$~\cite{Vij_MossbauerConversion_2006,LumsdenChristianson_Magnetism_FeHTS_JPCM_2010}.} \\
				\hline
				\NFA{} & 0.175(3) & 0.17(2)~\cite{TanSong_NaFeCoAs_Neutron_PRB_2016} &
				0.158(2)~\cite{PresniakovMorozov_111_Mossbauer_JPCM_2013}  \\ 
				BaFe$_2$As$_2$ & 0.75(5)~\cite{MallettYu_BaKFe2As2_EPL_2015} & 0.87(3)~\cite{HuangQiu_BaFe2As2_Neutron_PRL_2008} & 0.36(4)~\cite{RotterTegel_BaFe2As2_Mossbauer_PRB_2008} \\
				LaFeAsO~\footnote{Measured at $T
					\approx 25$ K, above the magnetic ordering temperature of La.} & 0.68(2)~\cite{deRenziBonfa_DFT_Muon_1111_SST_2012} &
				0.63(1)~\cite{QureshiDrees_1111_Neutron_Moment_PRB_2010} &
				0.34(1)~\cite{McGuireHermann_1111_Mossbauer_NJP_2009} \\ 
				FeSe$_{0.98}$ & 0.20(5)~\footnote{Taken under
					pressure $p =2.4$ GPa, from Ref.~\onlinecite{Bendele_PRB_2012}.} & Undetected~\footnote{No magnetic bragg peaks observed under pressure according to Ref.~\onlinecite{Bendele_PRB_2012}.} & 0.18(1)~\footnote{FeSe under pressure $p = 2.5$ GPa, from Ref.~\onlinecite{Kothapalli_11_Neutron_Moment_PRB_2010}.} \\
			\end{tabular}
		\end{ruledtabular}
		\label{tab:OrderedMomentComparison}
	\end{table}

	\section{Linear Relationship between Ordered Moment and \TN{}}
	\label{sec:Freq_TN_Linear}
	
	Despite the notable differences in the experimentally-measured ordered moment sizes across different
	Fe-HTS~\cite{LumsdenChristianson_Magnetism_FeHTS_JPCM_2010}, there are some notable relationships
	between the ordered moment and other material parameters. The observation of a linear
	relationship between the muon precession frequency $\nu$ and the magnetic ordering temperature \TN{}
	was initally noted by Uemura in Ref.~\cite{Uemura_PhysicaB_2009} for the ``122'' and ``1111'' classes of Fe-HTS.
	Separate linear trends in ``122'' and ``1111'' classes of Fe-HTS were discovered from M\"{o}ssbauer
	spectroscopy relating the internal hyperfine field and the orthorhombic lattice
	distortion~\cite{GoltzZinth_BaKFeCoAs2_MagSC_PRB_2014}. The different proportionality constants
	between the two classes of Fe-HTS have been ascribed to the critical dynamics of the structural and
	magnetic transitions~\cite{WilsonRotundu_Explain_Linear_PRB_2010,
		CanoCivelli_Explain_Linear_PRB_2010}. In this section, we make use of muon stopping site
	calculations to extend the investigation of the linear trend between the ordered moment size and
	\TN{} from \MuSR{} results.
	
	\subsection{Linear Trends from \MuSR{} Results}
	
	\begin{figure}[ht!]
		\centering
		\includegraphics[width=0.99\columnwidth]{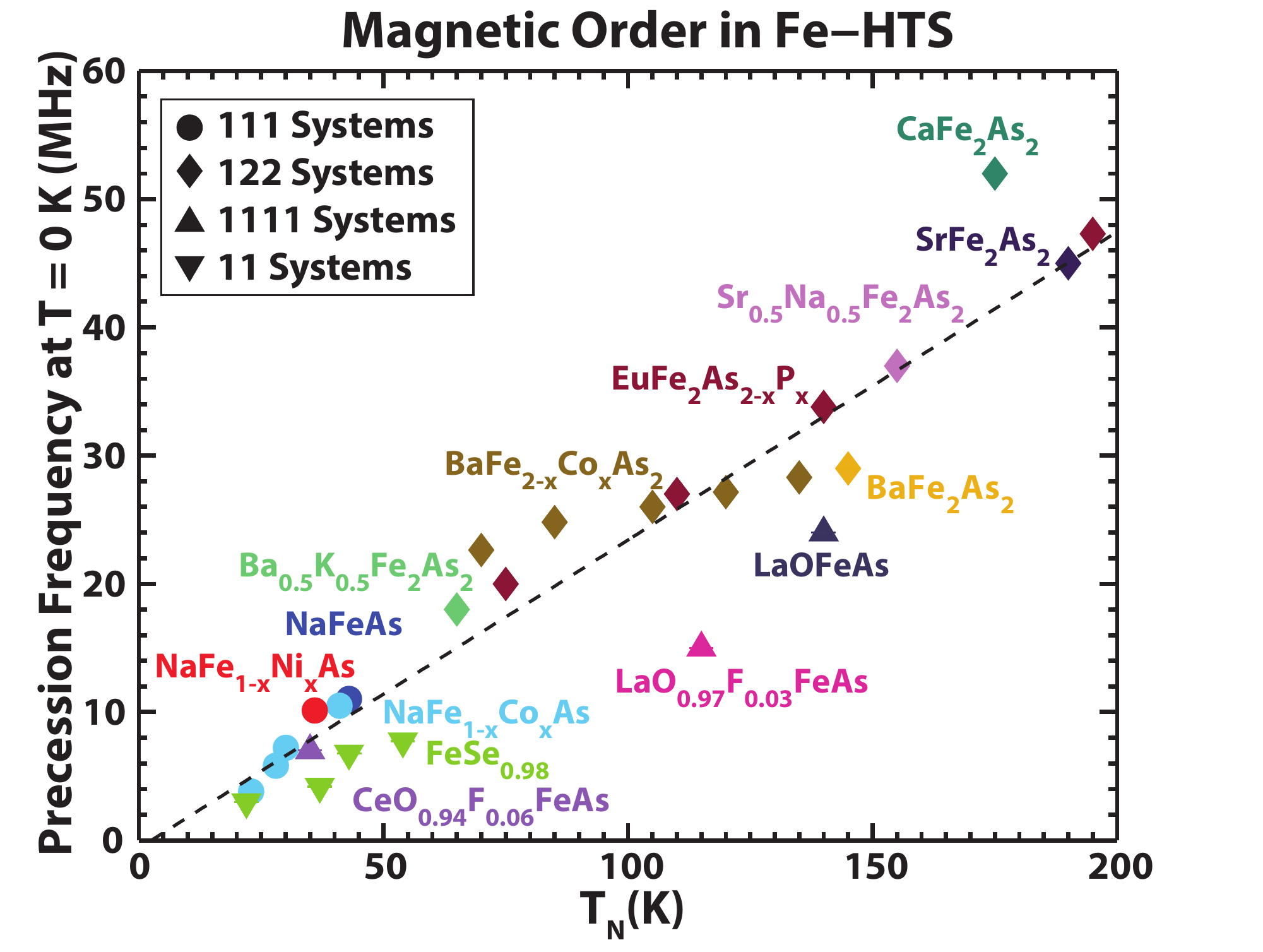}
		\linespread{1}\selectfont{}
		\caption{ 
			\label{fig:Frequency_TN_Linear}
			Correlation between low-temperature muon precession frequency $\nu(T\rightarrow 0)$ and the
			magnetic ordering temperature \TN{} for various Fe-HTS. The black dashed line is a linear
			least-squares model of the data. For systems with more than one precession frequency, the maximum
			frequency is taken. Circle symbols indicate the ``111'' family of Fe-HTS:
			\NFA{}~\cite{ParkerPitcher_MuSR_NaFeAs_CC_2009},
			\NFCoA{}~\cite{ParkerPitcher_MuSR_NaFeAs_CC_2009}, and NaFe$_{0.996}$Ni$_{0.004}$As. Diamond
			symbols indicate the ``122'' family of Fe-HTS: BaFe$_2$As$_2$~\cite{Uemura_PhysicaB_2009},
			Ba$_{0.5}$K$_{0.5}$Fe$_2$As$_2$~\cite{Aczel_PRB_2008},
			BaFe$_{2-x}$Co$_{x}$As$_{2}$~\cite{Bernhard_PRB_2012}, CaFe$_{2}$As$_{2}$~\cite{Goko_PRB_2009},
			EuFe$_{2}$As$_{2-x}$P$_{x}$~\cite{Guguchia_JSNM_2013}, SrFe$_{2}$As$_{2}$~\cite{Goko_PRB_2009},
			Sr$_{0.5}$Na$_{0.5}$Fe$_{2}$As$_{2}$~\cite{Goko_PRB_2009}. Upwards-pointing triangle symbols
			indicate the ``1111'' family of Fe-HTS: CaO$_{0.94}$F$_{0.06}$FeAs~\cite{Uemura_PhysicaB_2009},
			LaOFeAs~\cite{Luetkens_NatureMat_2009}, LaO$_{0.97}$F$_{0.03}$FeAs~\cite{Carlo_PRL_2009}.
			Downwards-pointing triangle symbols indicate the ``11'' family of Fe-HTS:
			FeSe$_{1-x}$~\cite{Bendele_PRB_2012} under pressure.
		}
	\end{figure}
	
	Figure~\ref{fig:Frequency_TN_Linear} depicts the correlation between the low temperature precession
	frequency $\nu(T \rightarrow 0)$ and the ordering temperature \TN{} for a variety of Fe-HTS,
	including \NFNA{} from the present investigation. Note that for the SC samples, the values of the precession frequencies, extrapolated to $T$ = 0 from
        above $T_{c}$ are taken. Since the precession frequency $\nu$ is
	proportional to the local magnetic field at the muon site, $\nu$ is proportional to the ordered
	moment size, and therefore the magnetization. $\nu$ also depends on the distance between the muon
	stopping site and the dominant ordered moment (Fe atoms). Remarkably, despite the differences in
	chemical composition and crystal structure across the various main families of Fe-HTS, (which
	influence the number and location of the muon stopping sites) a linear trend between $\nu(T
	\rightarrow 0)$ and \TN{} appears to persist. This suggests that the mechanism responsible for
	driving the magnetic ordering may be similar across different crystal structures and dopant atoms.
	Under this linear scaling relationship, there is an increase of 0.244(3)
	MHz/K between the $\nu(T \rightarrow 0)$ and \TN{}.
	
\begin{figure}[ht!]
		\centering
		\includegraphics[width=0.99\columnwidth]{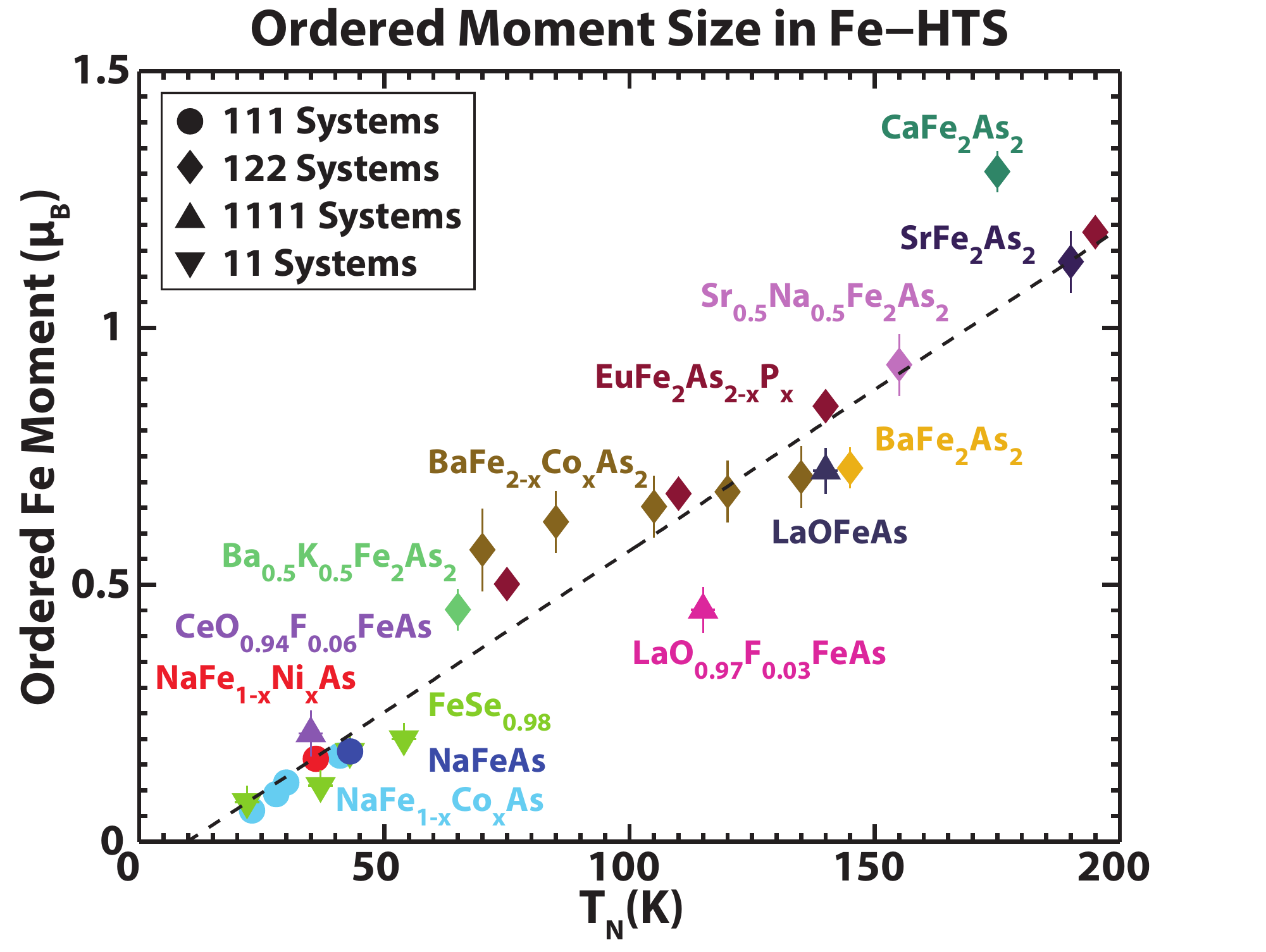}
		\linespread{1}\selectfont{}
		\caption{ 
			\label{fig:FeMoment_TN_Linear}
			Correlation between the low-temperature ordered moment size on the Fe atom $\mu_{\textrm{Fe}}$
			from \MuSR{} measurements and the magnetic ordering temperature \TN{} for various Fe-HTS. The
			black dashed line is a linear least-squares model of the data. The ``111'', ``122'', ``1111'', and
			``11'' families of Fe-HTS are represented by circle, diamond, upwards-pointing triangle, and
			downwards-pointing triangle symbols, respectively. See the caption for
			Figure~\ref{fig:Frequency_TN_Linear} for references to data points.
		}
	\end{figure}

	Since muon stopping sites have been calculated for a variety of Fe-HTS, we can compare the ordered
	Fe moment sizes directly. The ordered magnetic moment on the Fe atom can be calculated as the
	scaling factor necessary for matching the precession frequencies from dipolar field simulations
	against experimental results. Shown in Figure~\ref{fig:FeMoment_TN_Linear} is a comparison between the
	ordered magnetic moment of the Fe atoms and \TN{}. A linear model was fit to the data, revealing
	that the magnetic moment $\mu_{\textrm{Fe}}$ scales with \TN{} as 0.0062(6)
	$\mu_{\textrm{B}}/\textrm{K}$ across these families of Fe-HTS.
	
	\subsection{Discussion}
	
	The universal linear relationship between the $T = 0$ sub-lattice magnetization $M$ and the
	experimentally observed \TN{} provides important insight into the nature of the magnetic state.
	Within an itinerant mean-field approach, in which AFM is driven by Fermi surface nesting, $M \propto
	T_{\mathrm{N}}$ follows naturally whenever the Fermi surfaces are perfectly nested -- this is the
	same relationship between the SC gap function and \TC{} that appears in BCS theory. Since the Fermi
	pockets of the iron pnictides are not perfectly nested, it is important to verify whether $M \propto
	T_{\mathrm{N}}$ applies more generally in itinerant antiferromagnets. To investigate this issue, we
	consider a widely studied toy model consisting of one circular hole pocket located at the center of
	the Brillouin zone, and one elliptical electron pocket shifted from the center by the AFM ordering
	vector \cite{FernandesPratt_PairingFeSC_PRB_2010, KnolleEremin_Itinerant_PRB_2010,
		FernandesChubukov_Itinerant_PRB_2012, FernandesSchmalian_FreqTN_PRB_2010,
		VorontsovVavilov_FreqTN_PRB_2010,Almeida_PRB_2017}. The mismatch between the Fermi pockets is tuned
	by two parameters: $\delta_{2}$, which characterizes the ellipticity of the electron pocket, and
	$\delta_{0}$, which describes the difference between the areas of the Fermi pockets (and is
	therefore indirectly related to doping). The case $\delta_{0} = \delta_{2} = 0$ corresponds to
	perfect nesting, giving $T_{\mathrm{\mathrm{N},0}} = (\frac{e^{\gamma}}{\pi})M_{0} \approx 0.567
	M_{0}$.
	
\begin{figure}[ht!]
		\centering
		\includegraphics[width=0.99\columnwidth]{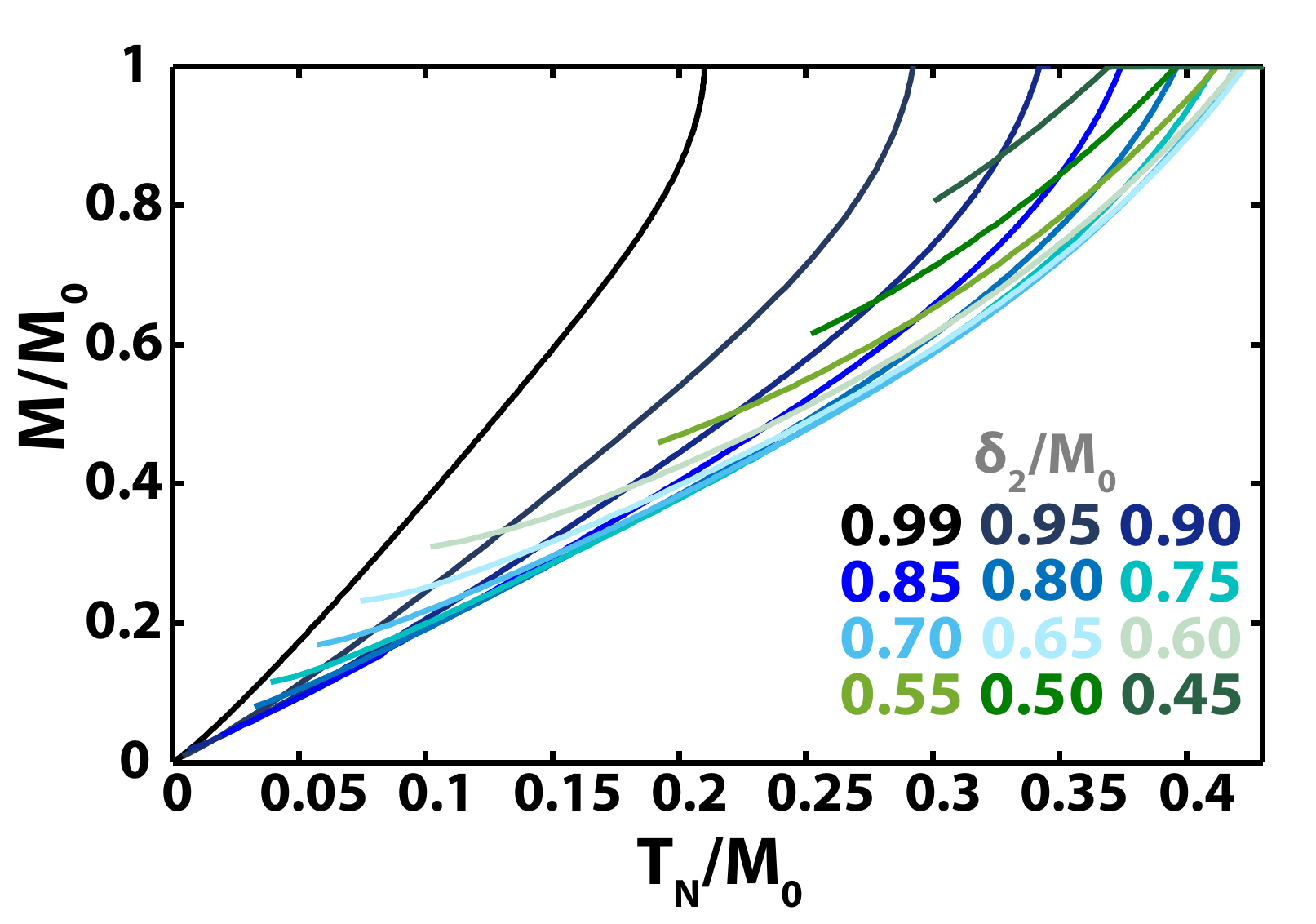}
		\caption{
			\label{fig:M_vs_TN}
			Theoretical calculations of the ordered moment $M$ at $T = 0$ versus the AFM critical temperature
			\TN{}. The ellipticity of the electron band, given by $\delta_{2}$, is fixed for each curve,
			whereas the parameter $\delta_0$, corresponding to doping, is varied continuously. $M_0$ is the
			AFM moment $M$ at $T=0$ when the hole and electron Fermi pockets are perfectly nested.
		}
	\end{figure}

	Following Refs.~\onlinecite{FernandesPratt_PairingFeSC_PRB_2010, FernandesSchmalian_FreqTN_PRB_2010,
		VorontsovVavilov_FreqTN_PRB_2010}, we compute not only \TN{} as a function of the parameters
	$\delta_{0}$ and $\delta_{2}$, but also the magnetization $M$ at $T = 0$. We focus on the regime in
	which the AFM transition is second order (see Appendix~\ref{sec:Theory_FreqTN} for details). As
	shown in Figure~\ref{fig:M_vs_TN}, $M$ monotonically increases with increasing \TN{}. Each curve
	corresponds to a fixed value of $\delta_{2}$ and continuously changing values of $\delta_{0}$.
	Interestingly, when $\delta_{2}$ is not too large, \TN{} and $M$ follow a nearly linear relationship
	over a wide parameter range, which is consistent with previous works
	\cite{FernandesPratt_PairingFeSC_PRB_2010, FernandesSchmalian_FreqTN_PRB_2010}. Although a
	quantitative comparison with experimental findings must account for band structure details of
	different compounds, the results of this simple model are qualitatively consistent with the
	experimental observations, suggesting that nesting plays an important role in driving the AFM
	instability.
	
	\section{Superconductivity in $\textrm{NaFe}_{1-x}\textrm{Ni}_{x}\textrm{As}$}
	\label{sec:TFMuSR}
	
	\MuSR{} experiments performed with an applied field transverse to the initial muon ensemble spin, called
	TF-\MuSR{}, allow determination of the magnetic field penetration depth $\lambda$, which is one of the
	fundamental parameters of a superconductor~\cite{Goko_PRB_2009}. (Recall that $\lambda$ is related
	to the superconducting carrier density $n_{s}$ through $\lambda^{-2}$ = $\mu_{0}e^{2}n_{s}/m^{\ast}$, where
	$m^{\ast}$ is the effective mass and $\mu_0$ is the vacuum permeability). Most importantly, the
	temperature dependence of $\lambda$ is particularly sensitive to the presence of SC nodes. In a
	fully gapped superconductor, $\Delta\lambda^{-2}\left(T\right) \equiv
	\lambda^{-2}\left(0\right)-\lambda^{-2}\left(T\right)$ vanishes exponentially at low $T$ and
	decays as a power of $T$ in a nodal SC. As a result, the \MuSR{} technique is a powerful tool to
	measure $\lambda$ in type II superconductors. Specifically, \MuSR{} experiments in the vortex
	state of a type II superconductor allow the determination of $\lambda$ in the bulk of the sample, in
	contrast to many techniques that probe $\lambda$ only near the surface.
	
	To understand the temperature evolution of $\lambda$, it is informative to study the symmetry and
	structure of the SC gap. Significant experimental and theoretical efforts have focused on this issue
	in Fe-HTS~\cite{Stewart_FeSC_RMP_2011, Scalapino_FeSC_RMP_2012}. However, there is no consensus on a
	universal gap structure and the relevance for the particular gap symmetry for Fe-HTS, which are the
	first non-cuprate materials exhibiting superconductivity at relatively high temperatures.
	
	In contrast to cuprates, where the SC gap symmetry is universal, the gap symmetry and/or structure
	of the Fe-HTS varies across different systems. For instance, nodeless isotropic gap distributions
	were observed in optimally doped
	Ba$_{1-x}$K$_{x}$Fe$_{2}$As$_{2}$~\cite{DingRichard_BaKFe2As2_TF_EPL_2008,
		KhasanovEvtushinsky_BaKFe2As2_TF_PRL_2009},
	Ba$_{1-x}$Rb$_{x}$Fe$_{2}$As$_{2}$~\cite{GuguchiaKhasanov_PRB_2016}, and
	BaFe$_{2-x}$Ni$_{x}$As$_{2}$~\cite{Abdel-HafiezZhang_BaFeNi2As2_TF_PRB_2015} as well as in
	BaFe$_{2-x}$Co$_{x}$As$_{2}$~\cite{TerashimaSekiba_BaFeCo2As2_TF_PNAS_2009},
	K$_{x}$Fe$_{2-y}$Se$_{2}$~\cite{ZhangYang_KFe2As2_TF_NatureMaterials_2011}, and
	FeTe$_{1-x}$Se$_{x}$~\cite{MiaoRichard_FeTeSe_TF_PRB_2012, BiswasBalakrishnan_FeTeSe_TF_PRB_2010}.
	Signatures of nodal SC gaps were reported in LaFePO~\cite{FletcherSerafin_LaFePO_TF_PRL_2009},
	LiFeP~\cite{HashimotoKasahara_LiFeP_LiFeAs_TF_PRL_2012},
	KFe$_{2}$As$_{2}$~\cite{DongZhou_KFe2As2_TF_PRL_2010},
	BaFe2(As$_{1-x}$P$_{x}$)$_{2}$~\cite{HashimotoYamashita_BaFe2AsP2_TF_PRB_2010,
		YamashitaSenshu_BaFe2AsP2_TF_PRB_2011, NakaiIye_BaFe2AsP2_TF_PRB_2010,
		ZhangYe_BaFe2AsP2_TF_NatPhys_2012},
	BaFe$_{2-x}$Ru$_{x}$As$_{2}$~\cite{QiuZhou_BaFeRu2As2_BaFe2AsP2_TF_PRX_2012}, and
	FeSe~\cite{SongWang_FeSe_TF_Science_2011} as well as in overdoped
	Ba$_{1-x}$K$_{x}$Fe$_{2}$As$_{2}$~\cite{KimStewart_BaKFe2As2_TF_PRB_2015} and in optimally doped
	Ba$_{1-x}$Rb$_{x}$Fe$_{2}$As$_{2}$ under pressure
	\cite{GuguchiaAmato_Pressure_BaRbFe2As2_TF_NatComm_2015}. Therefore, it is fruitful to extend the
	study of the SC gap symmetry to other Fe-based materials, specifically the ``111'' family of Fe-HTS.
	In this section, we present and discuss TF-\MuSR{} results on the $x = 1.3\%$ sample in the
	superconducting state.
	
	\subsection{TF-\MuSR{} Results}
	
	\begin{figure}[ht!]
		\centering
		\includegraphics[width=0.99\columnwidth]{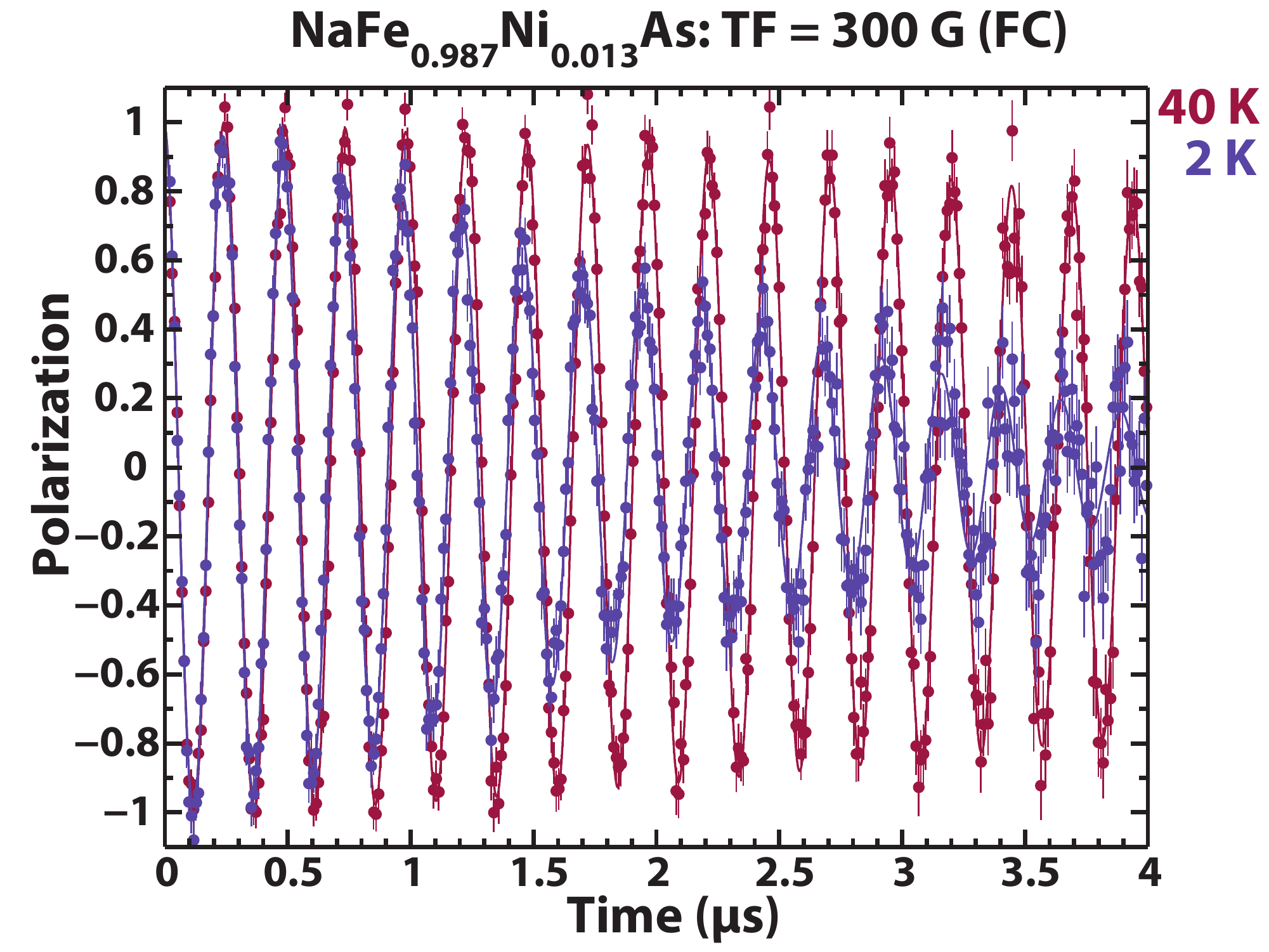}
		\linespread{1}\selectfont{}
		\caption{ 
			\label{fig:TF_Polarization}
			TF-\MuSR{} polarization on field-cooled $x = 1.3$\% in an applied field of 300 G. Time spectra in the $x = 1.3\%$ system with an applied transverse field at $T = 40$ K (PM) and $T = 2$ K (SC).
		}
	\end{figure}
	
	Shown in Figure~\ref{fig:TF_Polarization} are the TF-\MuSR{} time spectra on the $x = 1.3\%$ system,
	measured in an applied field of 300 Oe above (40 K) and below (2 K) \TC{} $\approx 15$ K. Above
	\TC{}, the oscillations show a small relaxation due to random local fields from nuclear magnetic
	moments. As the sample is field-cooled below \TC{}, the relaxation steadily increases due to the
	presence of a nonuniform local field distribution as a result of the formation of a flux-line
	lattice (FLL) in the SC state. The TF-\MuSR{} spectra were analyzed using the following functional
	form:
	
	\begin{figure}[ht!]
		\centering
		\includegraphics[width=0.98\columnwidth]{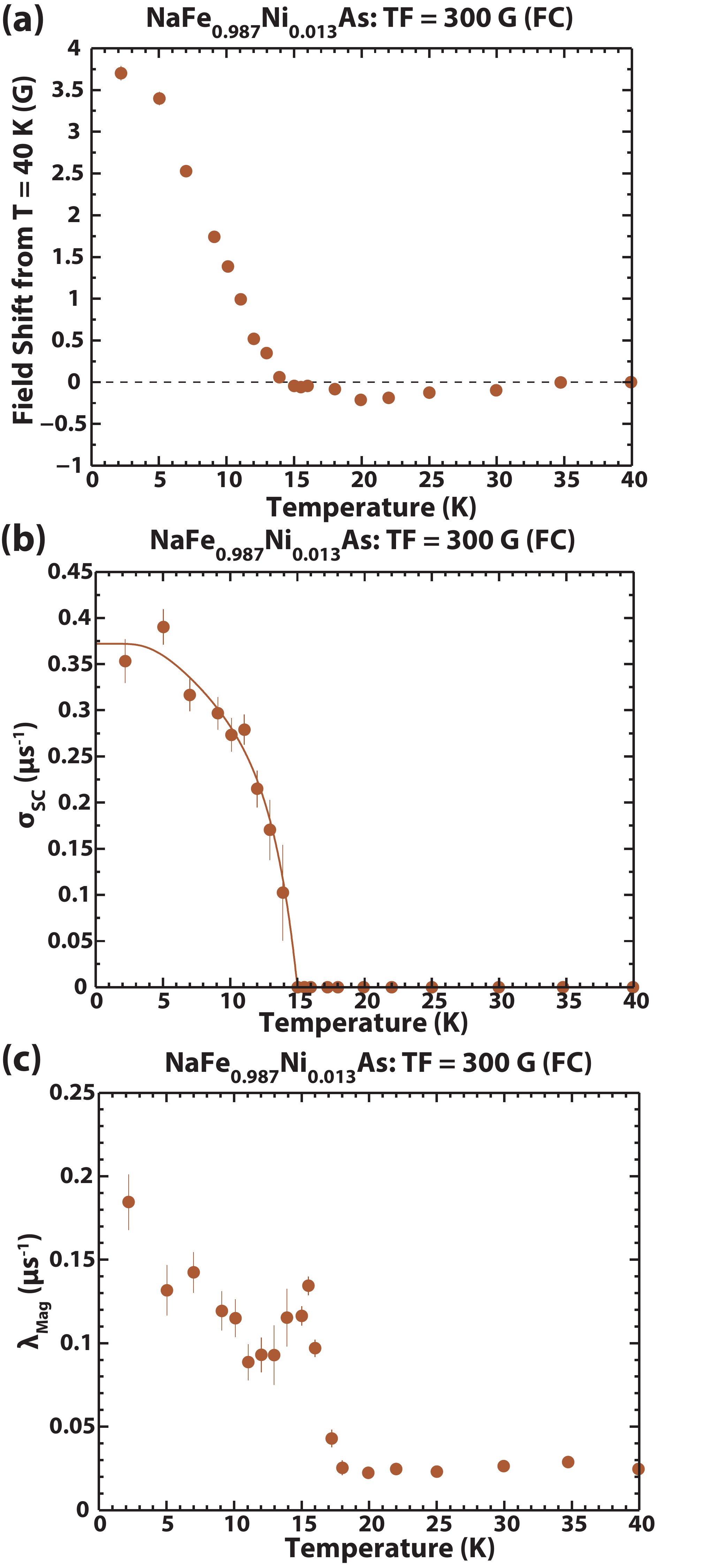}
		\linespread{1}\selectfont{}
		\caption{ 
			\label{fig:TF_FitParameters}
			TF-\MuSR{} results on field-cooled $x = 1.3$\% in an applied field of 300 G. (a) Temperature
			dependence of the field shift from 40 K (normal state). (b)-(c) Temperature dependences of the
			relaxation rates ascribed to the SC and magnetic orders, respectively. The solid line in (b)
			represents an isotropic two-band SC model fit to the temperature evolution of
			$\sigma_{\textrm{SC}}$. The peak in $\lambda_{\textrm{Mag}}$ in (c) is close to \TC{}, indicating
			that the onset of SC order affects the dilute electronic moment distribution.
		}
	\end{figure}
	
	\begin{align}
	\label{eq:TFPolarizationFit}
	\nonumber P_{\textrm{TF}}(t) &= F_{\textrm{nm}} \cos(2\pi\nu t + \phi)  \\
	&\qquad \times \exp\left\{ -\frac{1}{2} \left(\sigma^2_{\textrm{nm}} + \sigma^2_{\textrm{SC}} \right)t^2 \right\} e^{-\lambda_{\textrm{Mag}} t} 
	\end{align} 
	
	The defining parameters in~\eqref{eq:TFPolarizationFit} are the precession frequency $\nu$, the
	relaxation rates $\sigma_{\textrm{SC}}$ and $\sigma_{\textrm{nm}}$ characterizing the damping due to
	the formation of FLL in the SC state and the nuclear magnetic dipolar contribution, respectively,
	and an exponential relaxation rate for field-induced magnetism
	$\lambda_{\textrm{Mag}}$~\cite{Sonier_FieldInduced_PRL_2011}. The model
	in~\eqref{eq:TFPolarizationFit} has been previously used~\cite{Khasanov_FieldInduced_PRL_2009,
		GuguchiaKhasanov_PRB_2016} for Fe-HTS in the presence of dilute or fast fluctuating electronic
	moments and it was demonstrated to be sufficiently precise for extracting the SC depolarization rate
	as a function of temperature.
	
	The temperature dependence of $\nu$ shows a PM shift below \TC{} in
	Figure~\ref{fig:TF_FitParameters}(a), which is different from the expected diamagnetic shift imposed
	by the SC state~\cite{Khasanov_FieldInduced_PRL_2009, GuguchiaKhasanov_PRB_2016,
		Williams_FieldInduced_PRB_2010}. It is difficult to elucidate the origin of the PM shift, however
	the effects are consistent with field-induced magnetism. Other phenomena such as vortex lattice
	disorder~\cite{Sonier_FieldInduced_PRL_2011} or a Yosida-like decrease of the spin
	susceptibility~\cite{Yosida_SpinSC_PRL_1958} may also contribute to this behavior and can be
	investigated further.
	
	The SC and magnetic relaxation rates, $\sigma_{\textrm{SC}}$ and $\lambda_{\textrm{Mag}}$,
	respectively, are shown in Figure~\ref{fig:TF_FitParameters}(b)-(c), demonstrating an additional
	effect of a weak contribution of static magnetism to the SC state. We also observe the non-monotonic
	temperature dependence of $\lambda_{\textrm{Mag}}$, which may be caused by the interplay between
	magnetism and superconductivity~\cite{FernandesPratt_PairingFeSC_PRB_2010, Kang_MagSC_PRB_2015}. As
	the sample is cooled in an external transverse field below \TC{} $\approx 15$ K,
	$\sigma_{\textrm{SC}}$ begins to rise from 0 due to the FLL formation. $\sigma_{\textrm{SC}}$
	saturates upon further cooling, which resembles the behavior of an isotropic nodeless
	superconductor.
	
	We found that an isotropic two-band ($s$+$s$)-wave SC model describes the temperature dependence of
	the measured $\sigma_{\textrm{SC}}$ remarkably well (see Figure~\ref{fig:TF_FitParameters}(b)),
	yielding a large gap $\Delta_{1} \simeq 4.5(6)$ meV and a small gap $\Delta_{2} \simeq 1.8(5)$ meV.
	Refer to Appendix~\ref{sec:SCGap} for details on the SC gap symmetry analysis. A two-gap scenario is
	also consistent with the generally accepted view of multi-gap superconductivity in
	Fe-HTS~\cite{Stewart_FeSC_RMP_2011, HirschfeldKorshunov_SCGap_RPP_2011}. The magnitudes of the large
	$2\Delta_{1}/k_{\textrm{B}} T_{\textrm{C}} \simeq 6.9(5)$ and the small $2\Delta_{2}/k_{\textrm{B}}
	T_{\textrm{C}} \simeq 2.8(5)$ gap for \NFNA{} ($x$ = 0.013) are in good agreement with previous
	work~\cite{EvtushinskyInosov_SCGaps_NJP_2009}. There it was pointed out that most Fe-HTS exhibit a
	two-gap SC behavior, characterized by a large gap with magnitude $2\Delta/k_{\textrm{B}}
	T_{\textrm{C}} \simeq 7(2)$ and a small gap with $2.5(1.5)$.
	
	\subsection{Connection with Other Unconventional Superconductors}
	
	\begin{figure}[ht!]
		\centering
		\includegraphics[width=0.99\columnwidth]{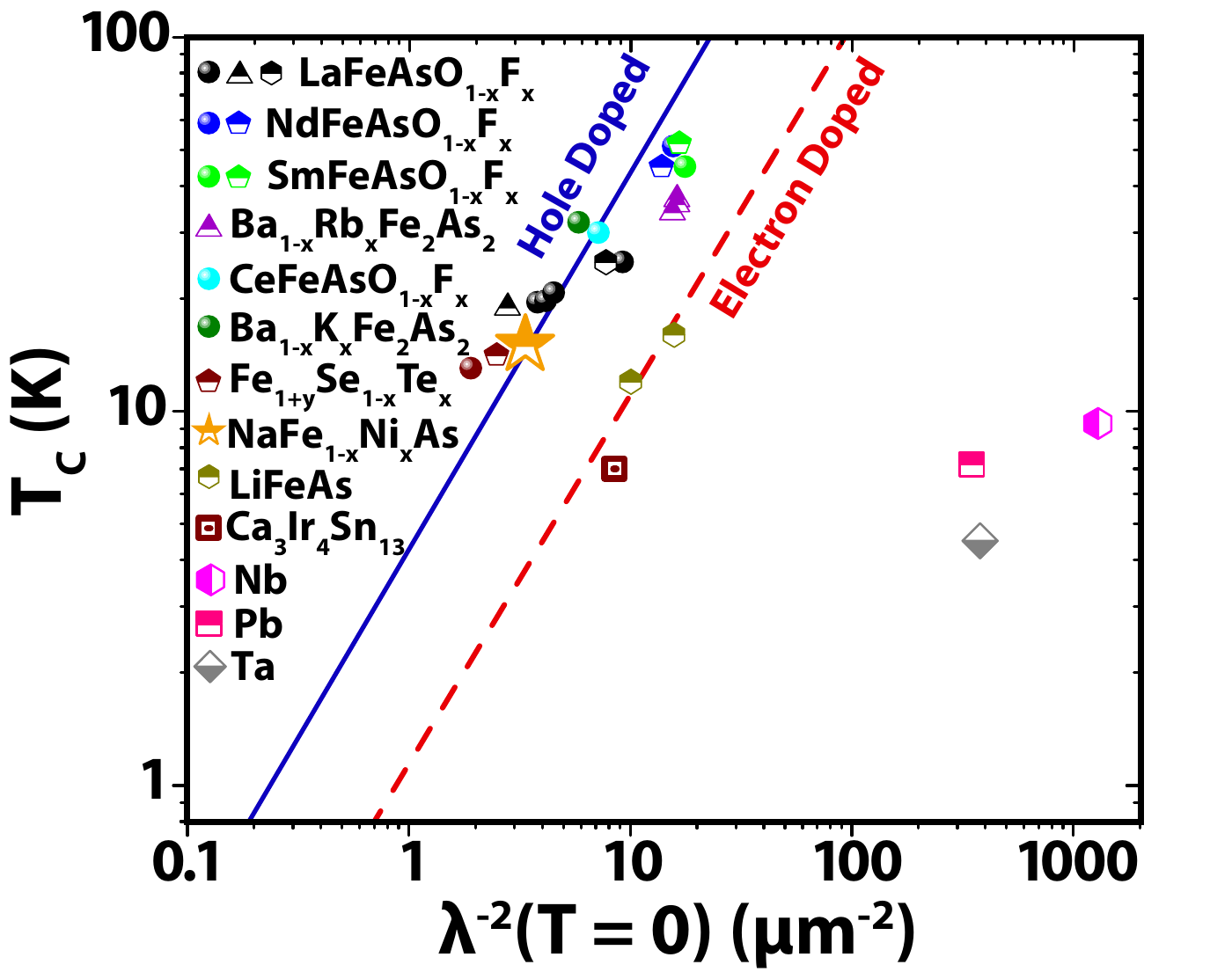}
		\linespread{1}\selectfont{}
		\caption{ 
			\label{fig:UemuraPlot}
			Uemura plot for hole and electron doped Fe-HTS (see Ref.~\onlinecite{Guguchia_UemuraPlot_NatComm_2017} and references therein).
			The linear relation observed for underdoped cuprates is shown as a blue colored solid line for hole
			doping~\cite{UemuraLuke_UemuraPlot_PRL_1989, UemuraKeren_UemuraPlot_Nature_1991} and as a red colored dashed
			line for electron doped systems~\cite{ShengelayaKhasanov_UemuraPlot_PRL_2005}. The points for
			conventional BCS superconductors are also shown. The orange star marker shows the data point for \NFNA{}
			obtained in this work.
		}
	\end{figure}
	
	An interesting result of \MuSR{} investigations in Fe-HTS is the observed proportionality between
	$T_{\rm c}$ and the zero-temperature relaxation rate $\sigma(0) \propto \lambda^{-2}(0)$, known as
	the Uemura plot~\cite{UemuraLuke_UemuraPlot_PRL_1989, UemuraKeren_UemuraPlot_Nature_1991}. This
	relation, which seems to be generic for various families of cuprate HTS, has the features that upon
	increasing the charge carrier doping $T_{\rm c}$ first increases linearly in the under-doped region
	(blue line in Figure~\ref{fig:UemuraPlot}), then saturates, and finally is suppressed for high
	carrier doping. The initial linear trend of the Uemura relation indicates that for these
	unconventional HTS, the ratio $T_{\textrm{C}}/E_{\textrm{F}}$ ($E_{\textrm{F}}$ is the Fermi energy)
	is much larger than that of conventional BCS superconductors. Figure~\ref{fig:UemuraPlot} shows
	\TC{} plotted against $\lambda^{-2}(0)$ for various hole- and electron-doped Fe-HTS (see
	Ref.~\onlinecite{Guguchia_UemuraPlot_NatComm_2017} and references therein), including the current
	results on \NFNA{}. The linear relation observed for underdoped cuprates is also shown as a solid
	line for hole doped cuprates~\cite{UemuraLuke_UemuraPlot_PRL_1989,
		UemuraKeren_UemuraPlot_Nature_1991} and as a dashed line for electron doped
	cuprates~\cite{ShengelayaKhasanov_UemuraPlot_PRL_2005}. The present data for \NFNA{} in the Uemura
	plot is in close proximity to the line observed in hole-doped cuprates and other Fe-HTS. This
	connection contrasts with \LFA{}, which shows behavior following electron-doped cuprates. The
	observation of a reduced superfluid stiffness in \NFNA{} compared to \LFA{} presents a new
	challenge for theoretical explanations.

	\section{Conclusion}
	\label{sec:Conclusion}
	
	In conclusion, the magnetic and SC properties of \NFNA{} were studied as a function of Ni-content
	$x$ by DC magnetization and \MuSR{} techniques. The long range magnetic order is observed for $x =
	0$ and 0.4\% samples, while for $x > 0.4\%$ magnetic order becomes inhomogeneous and is completely
	suppressed for $x = 1.5\%$. The magnetic volume fraction continuously decreases with increasing $x$.
	Furthermore, superconductivity acquires its full volume for samples with $x \gtrsim 0.4\%$. This
	implies that there is a coexistence of magnetism and superconductivity in \NFNA{}. Both the ordered
	moment and the magnetic volume fraction decrease below \TC{}, showing that magnetism, which develops
	at higher temperatures, becomes partially (or even fully) suppressed by the onset of
	superconductivity. These results indicate that the competition between the SC and magnetic order
	parameters in \NFNA{} develop in an intrinsically inhomogeneous environment, providing important
	insight for theoretical modeling. A linear relationship between the $T=0$ ordered moment and the AFM
	ordering temperature \TN{} for various Fe-HTS is noted, which is consistent with a mean-field
	approach for itinerant electrons, in which antiferromagnetism is driven by Fermi surface nesting.
	From TF-\MuSR{} measurements, the temperature evolution of the penetration depth in \NFNA{} is
	consistent with an isotropic twp-gap ($s$+$s$)-wave model for superconductivity.
	
	\begin{acknowledgments}
		
		The \MuSR{} experiments were performed at the Tri-University Meson Facility (TRIUMF) in Vancouver,
		Canada and at the Swiss Muon Source (S${\mu}$S) at Paul Scherrer Insitute (PSI) in Villigen,
		Switzerland. The authors sincerely thank the TRIUMF Center for Material and Molecular Science staff
		and the PSI Bulk \MuSR{} Group for invaluable technical support with \MuSR{} experiments. Work at
		the Department of Physics of Columbia University is supported by US NSF DMR-1436095 (DMREF) and NSF
		DMR-1610633. Z. Guguchia gratefully acknowledges the financial support by the Swiss National Science
		Foundation (SNF fellowships P2ZHP2-161980 and P300P2-177832). E.M. is supported by CNPq (grant number 304311/2010-3).
		P.B. acknowledges computing resources provided by STFC Scientific Computing Department's SCARF
		cluster. R.D.R. acknowledges funding by the European Unions Horizon 2020 research and innovation
		programme under grant agreement No 654000. This work was supported by the computational node hours
		granted from the Swiss National Supercomputing Centre (CSCS) under project ID sm07. R.M.F. is
		supported by the U.S. Department of Energy, Office of Science, Basic Energy Sciences, under Award
		number DE-SC0012336. C.D.C. acknowledges financial support by the National Natural Science
		Foundation of China Grant No. 51471135, the National Key Research and Development Program of China
		under contract No. 2016YFB1100101, and Shaanxi International Cooperation Program. Works at IOPCAS are supported by NSF and MOST of       China through Research Projects as well as by CAS External Cooperation Program of BIC (112111KYS820150017).  The present work
		is a part of the Ph.D. thesis of S.C.C. submitted to and defended at Columbia University in
		August 2017.
		
	\end{acknowledgments}
	

	\appendix

	\section{Internal Field Simulation}
	\label{sec:Computational}
	
	\subsection{Initialization of Crystal Properties}
	\label{sec:InitializationSimulation}
	
	At low temperatures, \NFA{} crystallizes into the Cmme space group, with the following assumed
	lattice constants for the orthorhombic structure based on
	Ref.~\onlinecite{LiCruz_NaFeAs_Neutron_PRB_2009}: $a = 5.6834$ \AA, $b = 5.6223$ \AA, and $c =
	6.9063$ \AA. Stopping site calculations and subsequent dipolar field calculations were performed on
	\NFNA{} with the atomic properties displayed in Table~\ref{tab:NaFeAsSetup}. The sample was assumed
	to be in the low temperature ordered state with Fe spins aligned in the usual colinear AFM
	arrangement as depicted in Figure~\ref{fig:Characterization}(d).
	
	\begin{table}[ht!]
		\centering
		\caption{
			Summary of crystal parameters for low-temperature simulations of \NFNA{}.
		}
		\begin{ruledtabular}
			\begin{tabular}{ccccc}
				Atom & Sym. & Position~\footnote{Atomic positions given in fractional coordinates.} & Nuc. Mom.~\footnote{Nuclear moments given in units of \muN{}.} & Mag. Mom.~\footnote{Ordered magnetic moments given in units of \muB{}.} \\
				\hline
				Na & 4g & (0.000, 0.250, 0.651) & 2.217 & -- \\ 
				Fe & 4a & (0.250, 0.000, 0.000) & 0.091 & 0.175 \\ 
				Ni & -- & Fe-substitution & -0.750 & -- \\
				As & 4g & (0.000, 0.250, 0.198) & 1.439 & -- \\ 
			\end{tabular}
		\end{ruledtabular}
		\label{tab:NaFeAsSetup}
	\end{table}
	
	\subsection{Muon Stopping Site Determination}
	\label{sec:DFT}

	
	The search for muon sites was initiated by sampling a $4 \times 4 \times 4$ grid of possible
	interstitial positions in the \NFA{} lattice that are at least 1 \AA{} away from lattice atoms.
	Symmetry-equivalent points in the search grid were removed with the spacegroup symmetry of the
	lattice. The stability of a H atom in a $2 \times 2 \times 2$ supercell consisting of 96 Na-Fe-As
	atoms was examined at each point in the grid. A $2 \times 2 \times 2$ Monkhorst-Pack grid of
	$\vec{k}$-points was used for Brillouin zone sampling. DFT calculations were carried out assuming
	the usual collinear magnetic ordering of Fe atoms in \NFA{} as shown in
	Figure~\ref{fig:Characterization}(d)~\cite{LiCruz_NaFeAs_Neutron_PRB_2009}. To accommodate for
	structural relaxations, the forces were optimized till a threshold of $10^{-3}$ atomic units and the
	energies till a threshold of $10^{-4}$ atomic units. Table~\ref{tab:MuonSiteDFT} lists five candidate
	muon sites for \NFA{} using this first-order search procedure. These sites are also assumed to be
	compatible for lowly doped \NFNA{}.
	
	\begin{table*}[ht!]
		\centering
		\caption{
			Summary of candidate muon stopping sites in \NFA{} determined by DFT. Muon site locations are
			believed to be similar for \NFNA{}. The local magnetic field strength $|\vec{B}|$ is nearly
			constant for each particular muon site position, confirming that the majority of the local field
			comes from the ordered Fe moments. The static ordered Fe moment was set to $\mu_{\textrm{Fe}} = 0.175(3)
			\mu_{\textrm{B}}$ to match the high frequency in the experimental spectra.
		}
		\label{tab:MuonSiteDFT}
		\begin{ruledtabular}
			\begin{tabular}{ cccccccc } 
				Cluster & Label & Symmetry & Site Position~\footnote{Candidate muon stopping site positions given in
					fractional coordinates.} & $\Delta E$ (meV)~\footnote{DFT total energy difference from stopping
					site A} & Field $|\vec{B}|$ (G)~\footnote{Magnetic field at muon site from dipolar field
					simulations} & Frequency $\nu$ (MHz)~\footnote{Simulated muon precession frequency} & Angle $\theta$ ($^\circ$)~\footnote{Average acute angle between the simulated field direction and the $c$-axis}\\ \hline
				I & A & 8n & (0.100, 0.750, 0.100) & 0 & 578.5(2.1) & 7.839(30) & 42.1(5)\\
				I & B & 8m & (0.000, 0.875, 0.100) & 42 & 810.9(3.5) & 10.987(49) & 31.1(6) \\
				I & C & 8l & (0.250, 0.500, 0.250) & 183 & 488.3(4.1) & 6.616(56) & 88.6(4)\\ \hline
				II & D & 4b & (0.750, 0.500, 0.500) & 287 & 1.002(69) & 0.014(40) & 0.2(3)\\
				II & E & 4g & (0.500, 0.250, 0.600) & 436 & 154.2(1.5) & 2.090(21) & 0.6(4)\\
			\end{tabular}
		\end{ruledtabular}
	\end{table*}
	
	\begin{figure}[ht!]
		\centering
		\includegraphics[width=0.99\columnwidth]{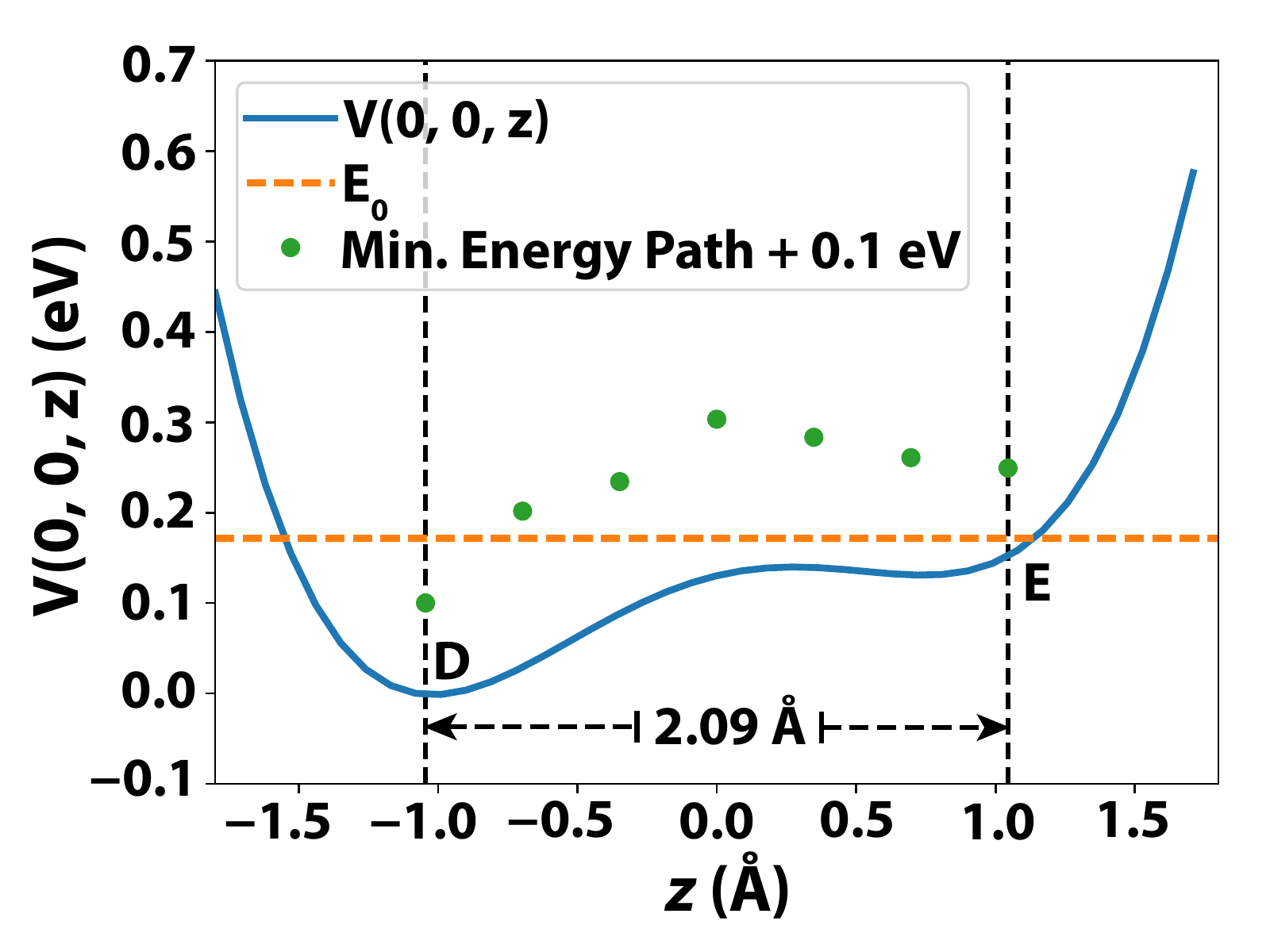}
		\linespread{1}\selectfont{}
		\caption{ 
			\label{fig:StoppingSiteStability}
			A toy model potential $V(0,0,z)$ (solid line) together with the ground state energy, $E_0 = 0.17$ eV
			from solving the Schr\"{o}dinger equation for a muon in a potential of the form $V(x,y,z) =
			\frac{1}{2}a(x^2 + y^2)+ \frac{1}{2}(bz^4-cz^2+dz) + f$ with $a = 2.44 \times 10^{-3}$, $b=5.04
			\times 10^{-4}$, $c = 3 \times 10^{-3}$, $d = 2.85 \times  10^{-3}$, and $f = 4.79 \times 10^{-3}$, all in
			Hartree atomic units. The green dots show the minimum energy profile map from the DBO for the
			symmetric site D to site E. These simulations imply that the muon is likely delocalized over the
			two sites in Cluster II (sites D and E).
		}
	\end{figure}
	
	We group the five candidate sites into two clusters based on stability checks using the Double
	Born-Oppenheimer approximation method (DBO)~\cite{BonfaSartori_DFT_Muon_2015}, which takes into
	account the quantum description of the muon. Within this method, a potential exploration algorithm
	(PEA) is used to efficiently sample the a priori unknown potential felt by the muon. With the
	sampling of the potential, site C is observed to be a local minimum in the muon potential. Site C
	relaxes towards sites A and B since site C has very low barrier less than 0.24 eV that is too small
	to bind the muon. Sites A and B are also close in proximity to each other and in energy difference.
	Consequently, we associate sites A, B, and C together as Cluster I. Similarly, we also observe that
	sites E and D relaxes into each other, which together form Cluster II. Our clustering also explains
	the observed frequencies - Cluster I contains the low DFT energy sites that describe the high muon
	field observed from experiment, while Cluster II contains sites that correspond to the low field.
	
	Shown in Figure~\ref{fig:StoppingSiteStability} are the results of further analysis of the sites in
	Cluster II. The energy profile extracted from the DBO potential map can be represented by the toy
	model shown in Figure~\ref{fig:StoppingSiteStability}. This enables us to solve the Schr\"{o}dinger
	equation of the muon, yielding a ground state energy of 0.17 eV (independent of the interpolation
	method and the boundary condition, to some extent) which is greater than the barrier seen in the
	potential map. These findings suggest that the muon wavefunction for the low field sites may be
	delocalized over positions between sites D and E (hereafter the D-E site). As a result, the low
	frequency detected from experiments may come from an averaging of the field at the two sites.
	Following analysis considering the quantum nature of the muon due to its light mass, we propose that
	sites A, B and D-E are the possible implantation sites of the muon.
	
	DBO would still predict a zero average at the D-E sites probed by the muon wavefunction due to the
	symmetry of the sites in the lattice. However a DFT mapping of the total energy and a separate
	solution of the muon Schr\"{o}dinger equation may not give the final answer, since the muon quantum
	nature is ignored in the DFT assessment of the total energy. The actual muon site may still be
	slightly distorting the local environment, thus justifying the small but nonvanishing low precession
	frequency listed in Table~\ref{tab:DFTDipolarSummary}.
	
	\subsection{Low Temperature Dipolar Field Simulation}
	\label{sec:DipolarSimulation}
	
	A $9 \times 9 \times 9$ supercell of magnetic dipoles was used to model the internal field of
	\NFNA{}. Dipole positions and strengths for the idealized crystal structure in \NFA{} listed in
	Table~\ref{tab:NaFeAsSetup}. Nuclear dipole moment directions are assumed to be random for all atoms
	while the spins on Fe are assumed to take on a collinear AFM striped pattern, common to other
	Fe-HTS. To simulate the effect of doping, the magnetic Fe atoms are randomly substituted with
	nonmagnetic Ni atoms to achieve the desired Ni concentration $x$. The dipolar field at the muon site
	was obtained by summing over all dipoles in the \NFNA{} supercell. 
	
	
	By comparing the simulated frequencies, shown in Table~\ref{tab:MuonSiteDFT}, with the experimental
	results, we can associate the two high frequencies $\nu_1$ and $\nu_2$ with sites B and A,
	respectively. The low frequency $\nu_3$ corresponds best with site E of Cluster II. However, our
	stability analysis shows that the muon is likely delocalized over sites D and E. A comparison
	between simulated and experimental results is presented in Table~\ref{tab:DFTDipolarSummary}. Our
	simulations show that the experimentally observed frequency $\nu_1 = 10.9$ MHz in \NFA{} corresponds
	to an ordered Fe moment size of about $\mu_{\textrm{Fe}} = 0.175(3)\mu_{\textrm{B}}$.

	\section{Ordered Moment Scaling Calculations}
	\label{sec:Theory_FreqTN}
	
	In this section, we present a description of the two-band model discussed in
	Section~\ref{sec:Freq_TN_Linear} and introduced in Refs.
	\onlinecite{VorontsovVavilov_FreqTN_PRB_2010, FernandesSchmalian_FreqTN_PRB_2010}. The effective
	free energy density of the model can be written as
	\begin{equation}
	\label{eq:FreeEnergyDensity}
	f = \frac{ 2 \mathcal{M}^{2} }{I} - \frac{T}{\upsilon} \sum_{\omega_{n}} \sum_{\boldsymbol{k}} 
	\ln [ ( \omega_{n}^{2} + E_{+,\boldsymbol{k}}^{2 } )( \omega_{n}^{2} + E_{-,\boldsymbol{k}}^{2} ) ] 
	\end{equation}
	where $\mathcal{M}$ is the temperature dependent ordered AFM moment, $I>0$ is the AFM interaction
	coupling constant, $\omega_{n} = 2 \pi T(n+1/2)$ is a fermionic Matsubara frequency
	($n\in\mathbb{Z}$), $\upsilon$ is the volume of the system, and
	
	\begin{equation*}
	E_{\pm,\boldsymbol{k}} = \sqrt{ \mathcal{M}^{2}+\xi_{k}^{2} } \pm |\delta_{\theta}| \mathrm{.}
	\end{equation*}
	Here, $\xi_{k}=k^{2}/2m-\epsilon_{0}$ is a parabolic energy dispersion, $\theta$ is the angle in the
	Fermi surface between the momentum $\boldsymbol{k}$ and the $x$-axis, and $\delta_{\theta} \equiv
	\delta_{0} + \delta_{2} \cos(2\theta)$ describes deviations from the perfect nesting condition.
	
	The momentum sum can be evaluated as $\frac{1}{\upsilon} \sum_{\boldsymbol{k}} \rightarrow
	\frac{m}{2\pi} \int_{-\infty}^{\infty} \diff\xi \int_{0}^{2\pi} \frac{\diff\theta}{2\pi}$. We
	minimize $f$ in~\eqref{eq:FreeEnergyDensity} with respect to $\mathcal{M}$ and perform the
	$\diff\xi$ integration to obtain
	\begin{equation}
	\label{eq:M_T_Finite}
	\frac{\mathcal{M}}{mI} = T \sum_{0 < \omega_{n} < \Lambda } \int\frac{\diff\theta}{2\pi} \Re \frac{\mathcal{M}}{\sqrt{\mathcal{M}^{2} + (\omega_{n} + i|\delta_{\theta}| )^2}} \mathrm{,}
	\end{equation}
	where $\Lambda$ is a high-frequency cutoff. At $T=0$, we integrated over frequencies to obtain
	\begin{equation}
	\label{eq:M_T_0}
	\Re \int\frac{\diff\theta}{2\pi} \ln \left( i|\delta_{\theta}| + \sqrt{ M^{2} - \delta_{\theta}^2 } \right) = \ln M_{0} \mathrm{,}
	\end{equation}
	where $M \equiv \mathcal{M}(T=0)$ and $M_{0}\equiv2\Lambda e^{-2\pi/mI}$ is the value of $M$ for
	$\delta_{0}=0=\delta_{2}$.
	
	When the transition is second order, \TN{} is the temperature at which $\mathcal{M}( T \rightarrow
	T_{ \mathrm{\mathrm{N}} } )\rightarrow 0 $. Setting $\mathcal{M}=0$ in~\eqref{eq:M_T_Finite} and
	performing the Matsubara sum yields
	\begin{equation}
		\label{eq:TN}
		2 \ln \left( \frac{ 4 \pi T_{\mathrm{\mathrm{N}} } } {M_{0}} \right) + \int \frac{\diff\theta}{2\pi} \left[ \psi\left(\frac{1}{2} + \frac{ i \delta_{\theta} }{ 2\pi T_{\mathrm{\mathrm{N}} } } \right) + \psi\left(\frac{1}{2} - \frac{ i \delta_{\theta} }{ 2\pi T_{\mathrm{\mathrm{N}} } } \right) \right] = 0
	\end{equation}
	where $\psi(z) \equiv \frac{\diff}{\diff z}\ln\Gamma(z)$ is the digamma function.
	
\begin{figure}
		\centering
		\includegraphics[width=0.99\columnwidth]{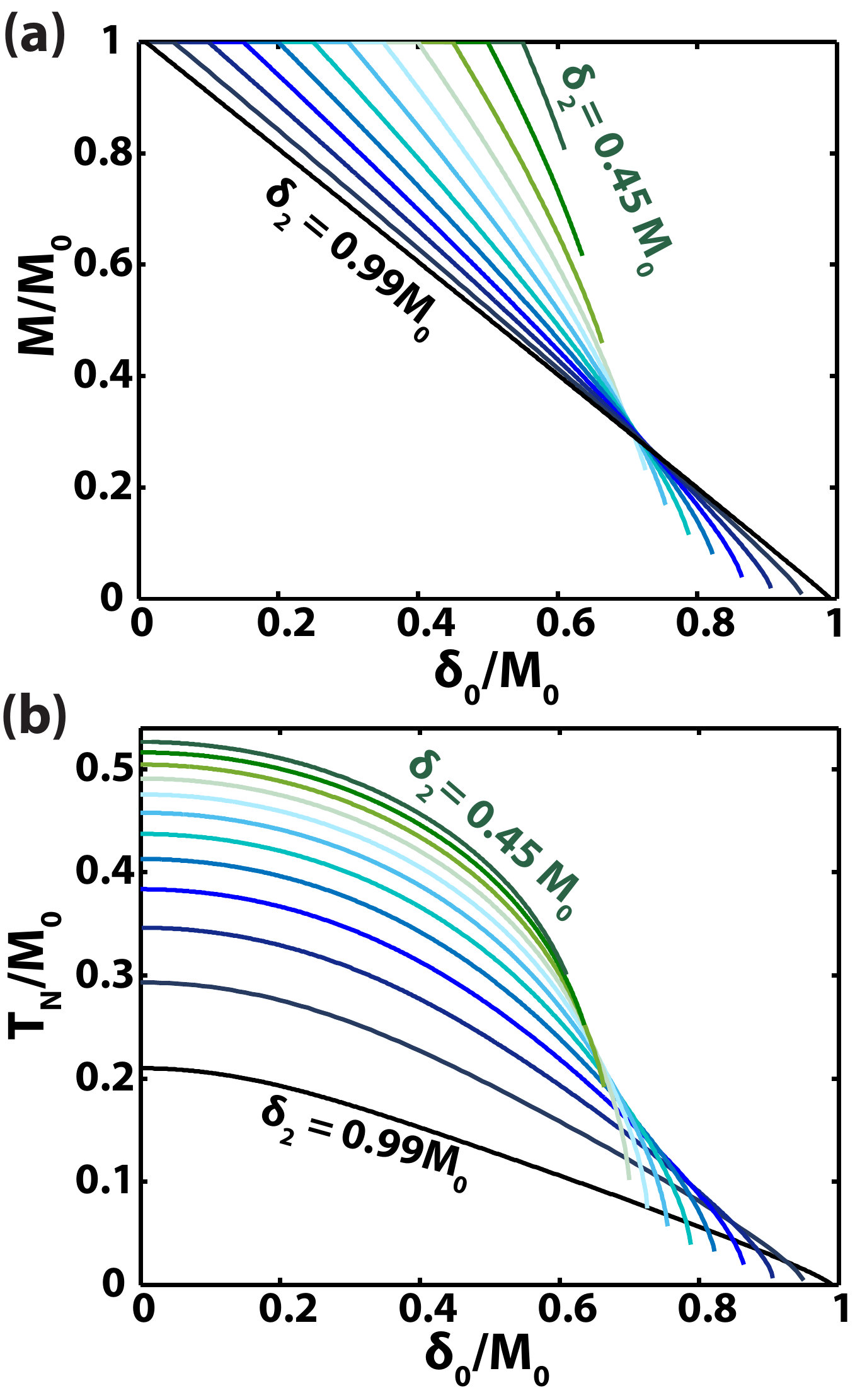}
		\caption{
			\label{fig:TN_vs_d0+M_vs_d0}
			Summary of numerical results for the ordered moment $M$ and \TN{}.
			(a) Ordered moment $M$ at $T=0$ and (b) \TN{} as functions of the area difference parameter
			$\delta_{0}$ for fixed values of the ellipticity parameter $\delta_{2}$. Results are shown in the
			regime where the transition is second order. Colors represent different values of $\delta_2 / M_0$
			listed in Figure~\ref{fig:M_vs_TN}.
		}
	\end{figure}

	We have numerically calculated the remaining angular integral in the self-consistent
	equations~\eqref{eq:M_T_0} and~\eqref{eq:TN} to determine the behavior of $M$ and \TN{} as functions
	of $\delta_{0}$ and $\delta_{2}$. For fixed $\delta_{2}$, the transition is second order at small
	$\delta_{0}$ but becomes first order at larger $\delta_{0}$. Moreover, there is no ordered AFM
	state~\cite{VorontsovVavilov_FreqTN_PRB_2010} for $\delta_{2}>M_{0}$. On the other hand, the ordered
	magnetic moment at $T=0$ generally increases with decreasing $\delta_{0}$ and $\delta_{2}$ and
	abruptly saturates at $\delta_{0} + \delta_{2} = M_0$. These results are presented in
	Fig.~\ref{fig:TN_vs_d0+M_vs_d0}. More importantly, these results enable us to plot and examine the
	behavior of the ordered moment at $T=0$ vs. \TN{}, which is shown in Fig.~\ref{fig:M_vs_TN}.
	
	\section{TF-\MuSR{} SC Gap Analysis}
	\label{sec:SCGap}
	
	To explore the SC gap symmetry, we recall that the penetration depth $\lambda(T)$ (in an isotropic
	superconductor) is related to the quadratic relaxation rate $\sigma_{\textrm{SC}}(T)$ through
	$\sigma_{\textrm{SC}}(T) = k \gamma_\mu \Phi_{0} \lambda^{-2}(T)$, where $\gamma_\mu$ is the muon
	gryomagnetic ratio, $\Phi_0 \equiv \frac{h}{2e}$ is the quantum of magnetic flux, and $k \approx
	0.06091$ is a geometric factor characterizing the FLL~\cite{MuSR:YaouancDalmas,
		Brandt_PenetrationSC_PRB_1988}. The temperature evolution of $\lambda(T)$ can be modeled for a
	variety of SC gap symmetries and structures.
	
	Within the local London limit of electrodynamics (where the penetration depth $\lambda$ is much
	greater than the SC coherence length $\xi$), the $\alpha$-model is a popular phenomenological
	framework used to study multiband superconductivity~\cite{Padamsee_AlphaSC_1973,
		DolgovKremer_AlphaSC_PRB_2005, Johnston_AlphaSC_Theory_2013}. The $\alpha$-model assumes that the SC
	gaps in different bands are independent from each other (aside from sharing a common \TC{}) and that
	the normalized penetration depth $\frac{\lambda(T)}{\lambda(0)}$ follows the same temperature
	dependence as in the single-band clean-limit BCS theory. A two-band $\alpha$-model in which the
	superfluid densities from each band are added together was used to analyze the TF-\MuSR{} results:
	\begin{equation}
	\label{eq:SC_2Gap_Alpha}
	\frac{\sigma_\textrm{SC}(T)}{\sigma_\textrm{SC}(0)} = \sum_{j= 1}^{2} f_j \frac{\lambda^{-2}(T,\tilde{\Delta}_{j})}{\lambda^{-2}(0,\tilde{\Delta}_{j})}
	\end{equation}
	where $\tilde{\Delta}_{j}$ is the maximum value of the SC gap at $T = 0$ for each band $(j = 1,2)$.
	The relative contributions from each band is imposed through the constraint $\sum_{j = 1}^2 f_j = 1$
	in~\eqref{eq:SC_2Gap_Alpha}.
	
	Assuming that the Fermi velocity is constant in magnitude, the penetration depth is determined
	through the integral expression~\cite{Carrington_PenetrationDepthSC_2003}:
	
	\begin{align}
	\label{eq:SC_PenetrationDepth}
	\nonumber \frac{\lambda^{-2}(T,\tilde{\Delta}_{j})}{\lambda^{-2}(0,\tilde{\Delta}_{j})} &= 1 +
	\frac{1}{\pi}\int_0^{2\pi}\diff \varphi \int_{\tilde{\Delta}_{j}}^{\infty} E \diff E  \times \cdots
	\\ &\quad \left(\pderiv{f}{E}\right) \frac{1}{\sqrt{E^2 - \Delta^2_j(T,\varphi)}}
	\end{align}
	where $f(E) \equiv [1 + \exp(E / k_{\textrm{B}}T) ]^{-1}$ is the Fermi function. The SC gap
	functions $\Delta_j(T,\varphi)$ in~\eqref{eq:SC_PenetrationDepth} are assumed to have the separable
	form: $\Delta_j(T,\varphi) = \tilde{\Delta}_{j} g(\frac{T}{T_{\textrm{C}}})S(\varphi)$. The
	temperature dependence of the gap is approximated by the function $g(t) = \tanh[\alpha (\beta
	(t^{-1} - 1)^\delta)]$, with $\alpha = 1.82$, $\beta = 1.018$, and $\delta =
	0.51$~\cite{Carrington_PenetrationDepthSC_2003}. The SC gap symmetry is embedded in $S(\varphi)$,
	which is defined to be 1 for $s$-wave and ($s$+$s$)-wave gaps and $|\cos(2\varphi)|$ for d-wave gaps.
	
	The results of applying~\eqref{eq:SC_2Gap_Alpha} to the $x = 1.3\%$ system are shown in
	Figure~\ref{fig:TF_FitParameters}(b), demonstrating that the two-band $\alpha$-model with an
	($s$+$s$)-wave SC gap is a feasible model for the data.

	\bibliography{NaFeNiAs_v21}	
\end{document}